\documentclass[aip,rsi,amsmath,amssymb,
reprint,
]{revtex4-1}

\usepackage{graphicx}
\usepackage{dcolumn}
\usepackage{bm}
\usepackage{amsmath}
\usepackage{eqnarray}
\usepackage{booktabs}
\usepackage{array}
\usepackage{makecell}
\usepackage{tabu}
\usepackage{epstopdf}
\usepackage[mathlines]{lineno}
\usepackage{color}

\begin{document}

\preprint{AIP/A160087}

\title[A novel facility for reduced-gravity testing]{A novel facility for reduced-gravity testing: a set-up for studying low-velocity collisions into granular surfaces}

\author{C. Sunday}
\altaffiliation[Currently employed at ]{the Jet Propulsion Laboratory, 4800 Oak Grove Drive, Pasadena, CA, 91107, USA}

\author{N. Murdoch}
\affiliation{D\'{e}partement Electronique, Optronique et Signal (DEOS), Syst\'{e}mes Spatiaux (SSPA), Institut Sup\'{e}rieur de l'A\'{e}ronautique et de l'Espace (ISAE-SUPAERO), Universit\'{e} de Toulouse, 31055 Toulouse, France}

\author{O. Cherrier}
\affiliation{D\'{e}partement M\'{e}canique des Structures et Mat\'{e}riaux (DMSM), Institut Sup\'{e}rieur de l'A\'{e}ronautique et de l'Espace (ISAE-SUPAERO), Universit\'{e} de Toulouse, 31055 Toulouse, France}
\affiliation{Institut Cl\'{e}ment Ader (CNRS UMR 5312), 31400 Toulouse, France}

\author{S. Morales Serrano}
\author{C. Valeria Nardi}
\author{T. Janin}
\author{I. Avila Martinez}
\affiliation{D\'{e}partement Electronique, Optronique et Signal (DEOS), Syst\'{e}mes Spatiaux (SSPA), Institut Sup\'{e}rieur de l'A\'{e}ronautique et de l'Espace (ISAE-SUPAERO), Universit\'{e} de Toulouse, 31055 Toulouse, France}

\author{Y. Gourinat}
\affiliation{D\'{e}partement M\'{e}canique des Structures et Mat\'{e}riaux (DMSM), Institut Sup\'{e}rieur de l'A\'{e}ronautique et de l'Espace (ISAE-SUPAERO), Universit\'{e} de Toulouse, 31055 Toulouse, France}
\affiliation{Institut Cl\'{e}ment Ader (CNRS UMR 5312), 31400 Toulouse, France}

\author{D. Mimoun}
\affiliation{D\'{e}partement Electronique, Optronique et Signal (DEOS), Syst\'{e}mes Spatiaux (SSPA), Institut Sup\'{e}rieur de l'A\'{e}ronautique et de l'Espace (ISAE-SUPAERO), Universit\'{e} de Toulouse, 31055 Toulouse, France}

\date{\today}

\begin{abstract}

This work presents an experimental design for studying low-velocity collisions into granular surfaces in low-gravity. In the experiment apparatus, reduced-gravity is simulated by releasing a free-falling projectile into a surface container with a downward acceleration less than that of Earth's gravity. The acceleration of the surface is controlled through the use of an Atwood machine, or a system of pulleys and counterweights. The starting height of the surface container and the initial separation distance between the projectile and surface are variable and chosen to accommodate collision velocities up to 20 cm/s and effective accelerations of $\sim$0.1 - 1.0 m/s\textsuperscript{2}. Accelerometers, placed on the surface container and inside the projectile, provide acceleration data, while high-speed cameras capture the collision and act as secondary data sources. The experiment is built into an existing 5.5 m drop-tower frame and requires the custom design of all components, including the projectile, surface sample container, release mechanism and deceleration system. Data from calibration tests verify the efficiency of the experiment's deceleration system and provide a quantitative understanding of the performance of the Atwood system.

\end{abstract}

                             

\maketitle
\section{Introduction}
\label{Introduction}

On November 12, 2014, the European Space Agency's Rosetta spacecraft became the first mission to successfully deliver a lander to a comet's surface. Though the Philae lander eventually came to rest, its anchoring harpoons failed to fire upon descent, and the lander proceeded to rebound twice over a duration of approximately 2 hours before reaching its final destination, roughly 1 km away from the intended landing site \cite{biele2015landing, Ulamec2015}. These unanticipated events led to significant changes in the operational schedule in order to perform as many scientific measurements as possible, within the limited lifetime of the lander.

The Rosetta events provide one example of how mission planning can be influenced by lander-surface interactions.  Hayabusa-2, a Japanese Space Agency (JAXA) asteroid sample-return mission, will be facing similar challenges to Rosetta in the coming years. The Hayabusa-2 spacecraft will arrive at the C-type near-Earth asteroid (162173) Ryugu in mid-2018 and deploy several science payloads to its surface \cite{Tsuda}. Among these payloads is a 10 kg lander, the Mobile Asteroid Surface Scout (MASCOT), provided by the German Space Agency (DLR) with cooperation from the French Centre National d'Etudes Spaciales (CNES). In addition to housing four instruments for in-situ science investigation, MASCOT contains a mobility mechanism that will correct its orientation and enable it to hop to various measurement sites \cite{Ho}. 

A lander similar to MASCOT is also proposed to be part of the European Space Agency's (ESA) Asteroid Impact Mission (AIM). AIM is one part of the Asteroid Impact \& Deflection Assessment (AIDA) mission together with NASA's Double Asteroid Redirection Test (DART) mission. AIM and DART complement each other in the validation of the kinetic impact approach to deflect threatening asteroids as well as in the characterization of physical and dynamical properties of the mission target, the Didymos binary asteroid system \cite{michelasteroid, cheng2015asteroid}.  

Based on thermal infrared observations \cite{harris2005surface, Campins, Gundlach, Moskovitz} and previous space missions \cite{sullivan1996geology, carr1994geology, thomas1999mathilde, robinson2002geology, Veverka, Yano, jaumann2012vesta}, it is strongly believed that asteroids are covered by loose regolith \citep[for more details see ][]{murdoch2015}. The asteroids' granular surfaces, in combination with the low surface gravity, make it difficult to predict a lander's collision behavior from existing theoretical models. This is partially due to the fact that granular materials have the ability to display either solid, liquid, or gaseous behavior. Anticipating rebound dynamics is particularly important for landers, like MASCOT, that do not have attitude control or propulsion systems. While an analysis of the Philae mission may assist in MASCOT's operation planning, further experimentation is required to construct and validate landing models.  

The objective of this work, derived from the needs of current and future small-body missions, is to present an experimental design for studying low-velocity collisions into granular surfaces in low gravity. Though this set-up is designed specifically around a collision study, this work shows how the experiment concept can be re-configured and used as a repeatable and inexpensive method for general reduced-gravity testing.

Section \ref{Background} outlines the main considerations behind the design of this experiment while Section \ref{Apparatus} describes set-up itself. Section \ref{Theory} explains the supporting theory used to construct the experiment set-up and Section \ref{Results} details its performance. Lastly, Sections \ref{Discussion} and \ref{Conclusions} discuss how this set-up will be used for future experimentation and how it can be reconfigured to accommodate other low gravity tests.

\section{Background on reduced gravity experimentation}
\label{Background}

Reduced-gravity experimentation has been a continued challenge for the Space industry. Present-day methods of simulating micro-gravity on Earth include parabolic flights, drop-towers, neutral buoyancy laboratories, magnetic levitation, rotating wall vessels, and off-loading gantries.

The budget and scheduling constraints associated with parabolic flights, large, air-evacuated drop towers, and neutral buoyancy laboratories can make these facilities impractical for frequent use. Furthermore, the majority of these methods are difficult to implement for more complicated, dynamic experiments. In order for neutral buoyancy or magnetic levitation to be effective for granular experiments, each particle must be neutrally buoyant or magnetically levitated. Off-loading weight from only the container or projectile will not suffice, because the particles inside of the container will still feel the effects of gravity.

Drop tower facilities and parabolic flights have been extensively used for micro-gravity experiments related to dust and regolith dynamics \cite{hofmeister2009flow, murdoch2013A, murdoch2013B, guttler2013, dove2013, Colwell2015}. Colwell and Taylor (1999) and Colwell (2003) studied micro-gravity collisions into granular surfaces over the course of two different payload experiments aboard the Space Transportation System (Space Shuttle) \cite{Colwell1999, Colwell2003}. Colwell et al. (2015) completed another series of micro-gravity impact tests over three parabolic flights, with the goal of studying low velocity collisions of centimeter sized particles  \cite{Colwell2015}. Goldman and Umbanhower (2008) observe reduced-gravity collisions into granular material by using an Atwood machine to change the effective gravity between a projectile and a surface at their moment of impact \cite{Goldman}. Altshuler et al. (2013) also employs an Atwood machine set-up in order to study extraterrestrial sink dynamics \cite{Altshuler}. An Atwood-type machine, using balanced weights linked in a loop, has also been used to simulate the landing of Huygens on Titan and for spacecraft impact tests on small bodies. These experiments had an emphasis on penetrometer tests and had impact velocities of 0.9 to 3 m/s \cite{Paton2015}. Other groups use drop-tower experiments to study impacts between solid and agglomerate bodies \cite{Beitz, Schrapler, Heibelmann}. By conducting their experiments aboard the Space Shuttle and NASA C-9 airplane, Colwell and Taylor (1999), Colwell (2003) and Colwell et al. (2015) are able to observe impact velocities ranging from 1 to 110 cm/s \cite{Colwell1999, Colwell2003, Colwell2015}. With drop-tower set-ups, Goldman and Umbanhower (2008) and Altshuler et al. (2013) observe collisions with higher impact velocities of 40 to 700 cm/s \cite{Altshuler, Goldman}. Altshuler et al. (2013) employs a setup similar to the one proposed in this work, but he uses polystyrene beads as a surface simulant \cite{Altshuler}. Additional studies are needed that combine the setup from Altshuler et al. (2013) with lower collision velocities and granular materials that are more representative of the regolith found on small bodies.

The key challenges to designing an asteroid collision experiment are: 1) Finding a way to simulate reduced gravity conditions on Earth, so that the prevailing forces in micro-gravity collisions can be reflected in the experimental results, and 2) simulating low-velocity collisions in such a way that data can be collected over a sufficiently-long time frame. Based on prior success and the opportunity for a customizable set-up, the proposed method to achieve this goal is through the use of an Atwood machine. In this approach, a free-falling projectile impacts a surface with a constant downward acceleration, or an acceleration less than that of gravity, so that the effective surface acceleration felt by the grains is very small. For example, if the projectile is in free-fall and the surface is controlled to have a downward acceleration of 9.0 m/s$^{2}$, then the surface experiences an effective acceleration of .81 m/s$^{2}$. In reducing the effective surface acceleration of the granular material, the confining pressure and interparticle friction will be reduced \cite{murdoch2013B} and the medium's inter-grain cohesion forces will become more important compared to its weight force \cite{Scheeres2010}. Consequently, the properties of the granular material will become more representative of those on an asteroid's surface. Since both the surface and projectile are falling in the Atwood machine set-up, the projectile requires a minimum amount of time to catch the surface before the collision begins. This extended free-fall period provides a solution to the second experimental design challenge and makes it possible to use accelerometers and high-speed cameras for data collection. This type of test would be impossible to perform on a parabolic flight because directional, low-levels of gravity are required to create low-velocity collisions with the granular surface. Gravity levels on a Zero-G aircraft tend to fluctuate around zero.

\begin{figure}
	\centering
	\includegraphics[scale=1.0]{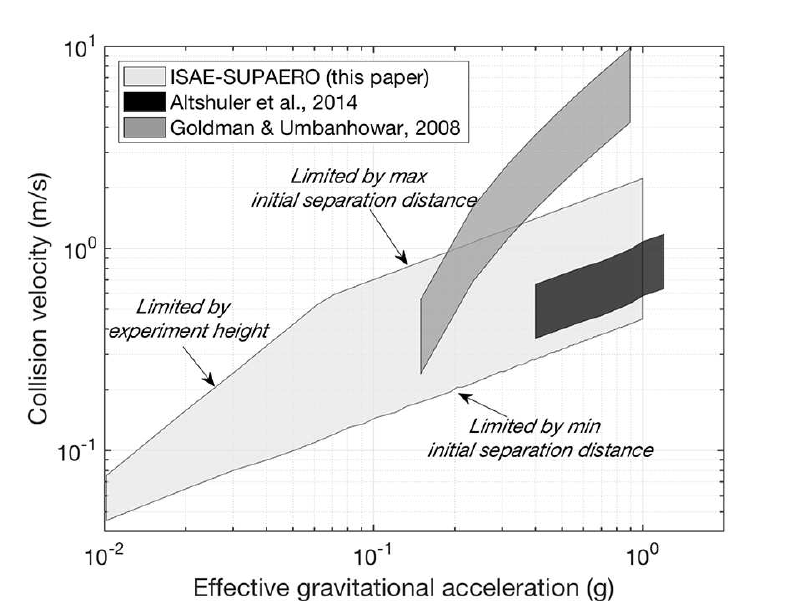}
	\caption[Relative regimes accessible from known Atwood machines]{Relative regimes accessible from known Atwood machines. Approximate values for the Asthuler et al. (2013) set-up are based on data extracted from Asthuler et al. (2013), Figures 2 and 3 \cite{Altshuler}. Approximate values for the Goldman and Umbanhower (2008) set-up are based on ranges of collision velocities and gravity levels provided in Goldman and Umbanhower (2008), Section II \cite{Goldman}. Theoretical values for the ISAE set-up are calculated as described in Section \ref{Theory}.}
	\label{fig:AtwoodMachines}
\end{figure}

Goldman and Umbanhower (2008) and Altshuler et al. (2013) conduct reduced-gravity testing using the same method as described above. However, the previous experiments are limited in the range of collision velocities and effective accelerations that their set-ups can attain. Figure \ref{fig:AtwoodMachines} compares the velocity and acceleration regimes that are studied by Goldman and Umbanhower (2008), Altshuler et al. (2013), and this work. The experiment in this work is built into an existing drop-tower structure, which is managed by the Department of Mechanical Structures and Materials at the Institut Sup\'{e}rieur de l'A\'{e}ronautique et de l'Espace (ISAE-Supaero) in Toulouse, France. With 5 meters of available drop height, 25 centimeters of possible separation between the projectile and surface, and a 160 kg surface container, the ISAE set-up can reach lower gravity levels and higher impact velocities than the set-ups used by Goldman and Umbanhower (2008) and Altshuler et al. (2013). The regime limits for the ISAE drop-tower are based on theoretical values and are dependent on the properties of the granular material being tested (see Section \ref{Theory}).

\section{Apparatus}
\label{Apparatus}

\begin{figure}
	\centering
	\includegraphics[scale=1]{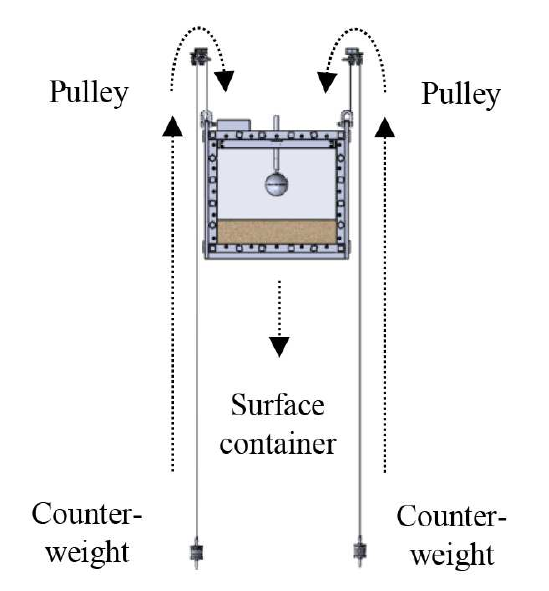}
	\caption[Functionality of an Atwood machine]{Isometric line drawing illustrating the basic operation of an Atwood machine.}
	\label{fig:AtwoodOperation}
\end{figure}

The idea of effective acceleration drives the design of this experiment and results in the following key features. First, the granular surface is provided a constant downward acceleration using an Atwood machine, or a system of pulleys and counterweights (see Figure \ref{fig:AtwoodOperation}). Next, the projectile and surface are simultaneously released from rest, where the starting height of the surface container and the initial separation distance between the projectile and surface are variable and selected based on the desired collision velocity and effective acceleration. Finally, at the end of the data collection period, the surface container is decelerated at a rate that does not result in damage to either the surface container or projectile.

\begin{figure}
	\centering
	\includegraphics[scale=1]{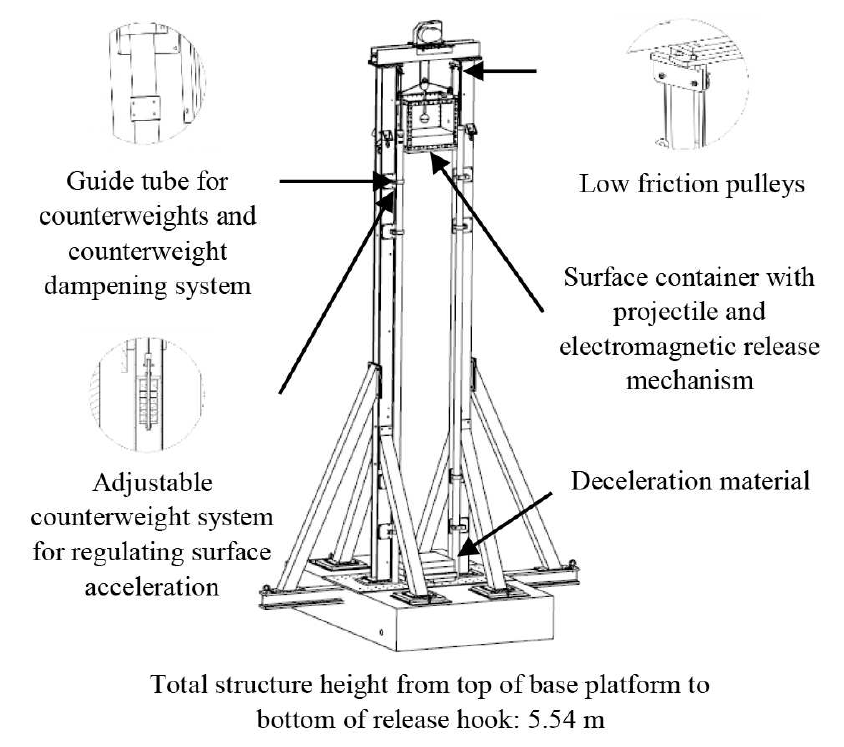}
	\caption[Isometric line drawing of the first experiment design]{Isometric line drawing of the experiment and existing drop tower structure.}
	\label{fig:fullAss}
\end{figure}

Figure \ref{fig:fullAss} shows a technical illustration of the full experiment, with reference to its six primary components: The support structure, the counterweight and pulley system, the surface container, the projectile, the release mechanism, and the deceleration system.

\subsection{Design of the drop tower base structure}

\begin{figure}
	\centering
	\includegraphics[scale=1]{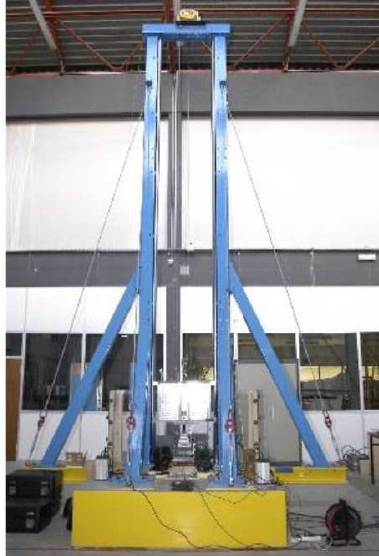}
	\caption[Tour de Chute]{ISAE-Supaero's existing drop tower structure. The tower features an electronically operated motor and pulley system for raising test materials and two low-friction rails for keeping the test materials vertically aligned.}
	\label{fig:tower}
\end{figure}

The experiment uses ISAE-Supaero's existing drop tower, shown in Figure \ref{fig:tower}, for structural support. The drop tower is usually designated for aircraft and material drop-tests \cite{israr2014}, but has been re-purposed to accommodate a counterweight and pulley system. This system makes it possible for materials to fall according to a predefined acceleration, and is key to performing reduced gravity, rather than microgravity, testing.

The tower provides approximately 5.5 x .65 x .65 meters of working space and includes an electronically operated motor and pulley system that can be used to raise the container assembly to the desired height. The tower also features two low-friction guide rails that permit test materials to fall and  have an impact normal to the structure's base.

\subsection{Design of the pulley system}

Four low-friction pulleys are mounted at the top of the structure's frame, and light-weight cords connect the surface container to the counterweights. The counterweights are enclosed inside a hollow tube so that their vertical acceleration and stopping motions can be controlled. The counterweight holders alone weigh 400 g each, and mass can be added to each holder at an increment of 100 to 250 g. The maximum counterweight mass that the current system can support, including the holder mass, is 8.8 kg. However, the holders can be replaced in order to accommodate larger masses. Figure \ref{fig:fullAss} shows the mounted pulley system, counterweight, and guide tube components of the assembly.

\subsection{Design of the surface container, release mechanism, and projectile}

\begin{figure}
	\centering
	\includegraphics[scale=1]{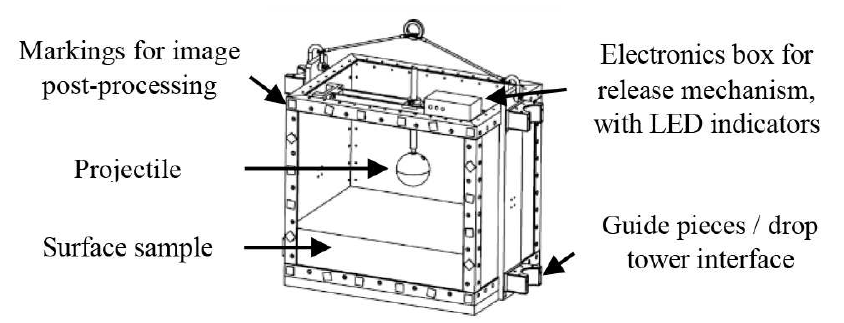}
	\caption[Isometric view the surface container]{Surface container sub-assembly, showing the 10 cm diameter projectile, the electromagnetic release mechanism, and reference markings for image post-processing. The surface container is 62 cm long, 45 cm wide, and 59 cm high.}
	\label{fig:container}
\end{figure}

The surface container sub-assembly comprises of three parts: the surface container, the release mechanism, and the projectile. The surface container is sized according to a literature review so that the walls of the container will not influence the rebound dynamics of the collision \cite{Goldman, Hartmann1978}. If the container is too small, then individual grains may interlock instead of moving in relation to one another as they would if unconstrained. The front and back panels of the surface container are made of 10 mm thick Makrolon polycarbonate material, while the two side panels are made of a light-weight 4 mm thick aluminum alloy. The container is fastened and reinforced at its joints using 2017 aluminum alloy members.  With the exception of the opening for the release mechanism, the container is closed on all four sides for safety reasons.  A narrow beam traverses the center of the container and acts as a support for the electromagnetic release mechanism. An electromagnet is mounted at the end of a supported tube, which can be raised and lowered to change the separation distance between the projectile and the surface.  The electronics box for controlling the electromagnet is mounted to the top of the container. Square markings of various orientation are placed on the front face of the container for use during the post-processing of the high-speed camera images. Lastly, two guide pieces fasten to the sides of the container and mate with the rails on the drop tower structure. These guide pieces constrain the container's motion to the vertical direction.  Figure \ref{fig:container} shows an illustration the surface container and the location of its different features. The total dry mass of the container assembly is 80 kg and it is filled to a height of approximately 17 cm with $\sim$80 kg of sand. Detailed discussion of the surface material used will be provided in Murdoch et al. (In Prep., 2016) \cite{Murdoch2016}.

The release mechanism consists of a permanent electromagnet and a contact switch. The contact switch is integrated into the release hook of the surface container such that the switch opens with the container's release. Using an onboard battery system and a electronic control card, the electromagnet is activated after the switch is opened. The activation of the electromagnet cancels the magnetic field of the permanent magnet allowing the projectile to fall. The time between the contact switch opening and the activation of the electromagnet can be adjusted via the onboard electronics to be between $\sim$40 and 400 ms. For these experiments, the shortest time (40 ms) was used. Testing confirms that the magnetic field generated by the electromagnet does not distort the sensors' measurements. An LED on the surface container's electronics box indicates when the contact switch is tripped so that the high speed camera images can be synchronized with the accelerometer data during post-processing.

\begin{figure}
	\centering
	\includegraphics[scale=1]{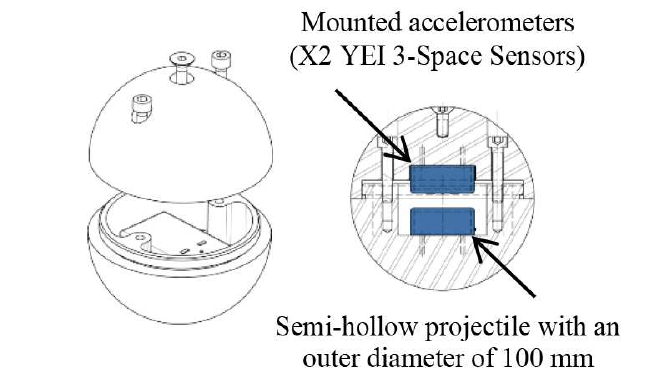}
	\caption[Projectile]{Semi-hollow experiment projectile, with two mounted YEI 3-Space Sensors}
	\label{fig:projectile}
\end{figure}

The experiment's projectile, shown in Figure \ref{fig:projectile}, is fabricated out of 2017 aluminum alloy and is designed to hold two wireless accelerometers (YEI 3-Space Sensors). Spherical markings of various radii are placed on the outside of the sphere and are used for target monitoring during the post-processing of the high-speed camera images. 

\subsection{Design of the deceleration system}
\label{decSection}

The surface container is decelerated using two 7 cm thick honeycomb panels. The HexWeb HRH-10 Aramid Fiber panels have a cell size of 4.8 mm and a density of 96 kg/m\textsuperscript{3}. Together, the panels can absorb at least 5 kJ of energy, or the equivalent energy of the falling surface container in the most extreme test scenario.

\subsection{Data acquisition}

In the experimental set-up, two different methods are available for capturing the motion of surface container and projectile. First, YEI 3-Space Sensors are mounted to the projectile and surface container. These sensors are data-logging devices that contain an Attitude and Heading Reference System (AHRS), an Inertial Measurement Unit (IMU), and a micro-SD card for on-board data storage. The YEI 3-Space Sensors were specifically selected because of their low mass (28 g), low dynamic range (selectable from $\pm2g$/$\pm4g$/$\pm8g$ with a noise density of 99$\mu g/\sqrt Hz$) and high shock resistance (up to 5000$g$)\cite{YEI}. These features allow the sensors to record the impact between the projectile and the sand with high precision and to survive to the final shock at the end of the drop. The sensors can also be integrated into the projectile without impacting its desired design features, such as size and weight. In addition to the accelerometers, an Ultima APX-RS Photron FASTCAM is used, with a Sigma 24-70mm f/2.8 DG lens, to capture high-speed images of the projectile collision at 1,000 frames per second and a 1024 x 1024 pixel resolution. The camera was placed at a distance of $\sim$2.7 m from front of the surface container, and a focal length of 24 mm was used for the lens.

\section{Theory}
\label{Theory}

In order to customize this novel Atwood set-up for different experimental trials, several attributes of the system must be calculated. For example, the available ranges of counterweight masses, separation distances between the projectile and container, and starting heights of the container must be identified. The following sections provide a theoretical understanding of the system and the scientific trials that are consequentially available given the set-up's physical constraints. 

\subsection{Sizing of the Atwood machine counterweights}

The ``reduced-gravity" element of the experiment is introduced through the use of a counterweight and pulley system, which allows the surface container to have a constant downward acceleration less than that of gravity.  If pulley friction and chord elasticity are neglected, then the controlled acceleration is simply a function of mass. The expression for the surface container's acceleration $a_s$ is derived by balancing the forces on the container (subscript, $s$) and counterweights (subscript, $cw$) and is given by Equation \ref{eq:cwMass}, where $m_s$ is the mass of the surface container and $m_{cw}$ is the total combined mass of all counterweights.  

\begin{equation} 
	\centering
	a_{s} = g\left(\frac{m_{s} - m_{cw}}{m_{s} + m_{cw}}\right)
	\label{eq:cwMass}
\end{equation}

\subsection{Calculating the required starting height for experimental trials}

It is necessary to calculate the starting drop height of each experimental trial so that a prediction can be made as to where the collision takes place along the height of the drop tower. Having this information serves two purposes. First, it helps bound what effective accelerations and collision velocities can be achieved within the physical limits of the drop tower, and second, it provides an estimate for where the high-speed cameras should be positioned in order to record the collision and rebound phases of the test.

The total drop height can be described by four phases experienced by the projectile: a free-fall phase, a collision phase, a rebound phase, and a deceleration phase. The freefall phase of the experiment begins when the surface (subscript, $s$) and projectile (subscript, $p$) are released from rest and ends when the two objects make contact for the first time. The collision phase begins at the end of the free-fall phase and ends when the projectile returns to the same position relative to the surface, after experiencing some level of surface deformation. Even though the projectile may never return to its starting position, such as in the case of a purely inelastic collision, the definition of this phase assumes it does in order to size for the most extreme height case. The rebound phase begins at the end of the collision phase and ends shortly before the projectile impacts the surface for a second time. The duration of the rebound phase depends on the collision's coefficient of restitution (COR), or ratio of energy lost during the collision. For the purposes of this calculation, a value for the collision's COR is estimated from a literature review (see Section \ref{estParameters}), and the rebound phase is considered to begin at the end of the collision phase and end when the projectile has reached a maximum theoretical separation from the surface. Using this alternative definition for the rebound phase decreases the total height of the experiment while still providing an adequate window of time to observe the projectile's rebound behavior. Finally, the deceleration phase begins at the end of the rebound phase and ends as soon as the surface container is brought to a complete stop.

The theoretical distance traveled and time elapsed during each phase can be derived from kinematic equations and energy conservation. The total height ($H_{total}$) required for each trial is then equal to the sum of the heights for the free-fall, collision, rebound and deceleration phases ($h_{sd}$, $h_{sc}$, $h_{sr}$ and $h_{sdec}$ respectively): 

\begin{equation} 
	\centering
	H_{total}= h_{sd}+h_{sc}+ h_{sr}+h_{sdec} 
	\label{eq:fullHeight1}
\end{equation}

The total height as a function of velocity and time is given by Equation \ref{eq:fullHeight}, where $V_c$ is the collision velocity, $a_s$ is the acceleration of the surface, $g$ is the acceleration of gravity, $V_{sc0}$ is the velocity of the surface container at the beginning of the collision phase, $t_{c}$ is the estimated collision time, $V_{scF}$ is the velocity of the surface container at the end of the collision phase, and $t_{r}$ is the estimated rebound time :

\begin{eqnarray}
	\centering
	H_{total} & = & \frac{V_c^2}{2}\left(\frac{a_s}{g^2 - 2a_{s}g + a_s^2}\right) + V_{sc0}t_{c} \nonumber \\
	& & +  \frac{1}{2}a_{s}t_c^2 + V_{scF}t_{r} +  \frac{1}{2}a_{s}t_r^2 + h_{sdec}\
	\label{eq:fullHeight}
\end{eqnarray}

\subsection{Estimated parameters}
\label{estParameters}

Two unknown parameters are required to calculate the total drop height for each experimental trial: the time that elapses during the collision phase ($t_c$) and the collision's coefficient of restitution (COR) that will determine the rebound time ($t_r$).

Different collision models can be used to predict the collision time, but only when certain properties of the medium are known. For example, if the equivalent stiffness $k^*$ and damping $c^*$ of the sand and projectile under low-gravity conditions were known, then the collision time $t_c$  could be estimated using the spring-dashpot collision model shown in Equation \ref{eq:springDash}, where $\omega _d$ is the damped natural frequency and $m_i$ is the mass of the impactor \cite{Nagurka}. 

\begin{equation} 
	\centering
	t_c = \frac{\pi}{\omega _d} = \frac{2\pi m_i}{\sqrt{4k^*m_i - c^{*2}}}
	\label{eq:springDash}
\end{equation}

When the Young's modulus $E$ of the surface material is known, $t_c$ can be estimated using Hertz collision theory for the elastic collision of two spheres:

\begin{equation} 
	\centering
	t_c = 2.87 \left(\frac{m^{*2}}{RE^{*2} V_c}\right)^{\frac{1}{5}}
	\label{Hertz}
\end{equation}

\noindent where $E^*$ is the equivalent Young's modulus of the impactor and surface materials, $m^{*}$ is the equivalent mass of the two spheres, and $R$ is the reduced radius of the two spheres \cite{Krijt}:

\begin{align} 
	\frac{1}{E^*} = \frac{1-{\nu_1}^2}{E_1} +  \frac{1-{\nu_2}^2}{E_2}; \\
	\frac{1}{R} = \frac{1}{R_1} +  \frac{1}{R_2}; \\
	\frac{1}{m^*} = \frac{1}{m_1} +  \frac{1}{m_2};
\end{align}

\noindent While the Hertz collision model does not accurately represent collisions into granular materials, it can still be used to obtain a lower bound estimate on collision time. For example, using the Hertz collision model, Krijt et al. (2013) predicts a collision time of 1.4 x $10^{-8}$ s for an 8 m/s head-on collision between icy spheres \cite{Krijt}. This is the lower-bound collision-time estimate because the Hertzian model considers solid-to-solid body collisions, as opposed to much more inelastic solid-to-granular surface collisions \cite{Goldman, Krijt}.

Collision times observed in previous granular impact experiments range from 0.04 seconds to 0.5 seconds \cite{Goldman, Ambroso2005, Altshuler}. Altshuler et al. (2013) observes the longest collision times of 0.2 - 0.5 seconds while studying the sink dynamics of a 23 g sphere into non-cohesive polystyrene beads \cite{Altshuler}. Goldman and Umbanhower (2008) find a collision time of 0.1 seconds for a 10 g disk impacting glass beads at 60 cm/s, a collision time of 0.08 seconds for a steel sphere with a 1.91 cm radius impacting glass beads at 2.86 m/s, and a collision time of 0.07 seconds for a 147 g nylon sphere impacting glass beads at 47 cm/s \cite{Goldman}. Goldman and Umbanhower (2008) conclude that the collision time for a steel sphere colliding with glass beads is independent on impact velocity for sufficiently high impact velocities, but that a regime change occurs for low collision velocities ($\lessapprox$ 1.5 m/s), where collision time actually increases with decreasing impact velocity \cite{Goldman}. A study by Ambroso et al. (2005) finds a collision time of approximately 0.04 seconds for a 5 cm diameter wooden sphere impacting glass beads at 226 cm/s \cite{Ambroso2005}.

Since the properties of sand under low gravity conditions are unknown and the collisions are likely to be inelastic, it is difficult to predict the collision time using the Spring-dashpot and Hertz collision models. In order to size the experiment structure, an estimate of 0.1 seconds is used for the collision time. This estimate is based on the study by Goldman and Umbanhower (2008). Though results from Altshuler et al. show longer collision times, Goldman and Umbanhower use a more comparable experimental set-up and provides an estimate that still allows for a conservative collision-height calculation.

The second unknown parameter in the height equation is the collision's coefficient of restitution. Space Shuttle payload experiments by Colwell and Taylor (1999) and Colwell (2003) show that the COR for impacts at less than 12 cm/s are either unobservable or as low as 0.02 \cite{Colwell1999, Colwell2003}. At the other extreme, analysis of the Hayabusa-1 touchdown data on asteroid Itokawa suggests that the COR is as large as 0.84 \cite{Yano}, although their are many uncertainties associated with this measurement. To make an approximation on the height calculation, an average value for the COR is estimated based on lower and upper bound observations from previous impact experiments. The average value is used instead of the highest value because the beginning of the rebound phase can still be observed even if the COR is underestimated.

The lower estimate is taken as 0.02 based on Colwell (2003), and the upper estimate on COR is taken from studies observing low-velocity collisions between centimeter-sized dust aggregates in drop-tower configurations. Beitz et al. (2011) finds an average COR of 0.35 $\pm$ 0.12 for collision velocities of 0.8 to 37 cm/s between agglomerates made of SiO$_{2}$ micrometer-sized dust particles \cite{Beitz}. In another drop-tower experiment that extends the work of Beitz et al. (2011), Schr{\"a}pler et al. (2012) develops an expression for COR as a function of impact velocity for collision velocities in the range of 1-10 cm/s \cite{Schrapler}. This expression, shown in Equation (\ref{eq:schraplerCOR}), acts as the upper-bound for the COR estimate.

\begin{equation} 
	\centering
	COR_{max} = .11 V_c^{-0.51} {(m/s)}^{0.51} 
	\label{eq:schraplerCOR}
\end{equation}

Based on the studies of Colwell, Bietz et al., and Schr{\"a}pler et al., the COR of the collision is estimated using Equation \ref{eq:COR} for collision velocities $>$1 cm/s. This estimate is not the expected COR of the collision, but simply a reasonable guess that will allow for the sizing of the experiment. 

\begin{equation} 
	\centering
	COR_{mean} = \dfrac{0.02 + .11 V_c^{-0.51}}{2} {(m/s)}^{0.51} 
	\label{eq:COR}
\end{equation}

\subsection{Tower capabilities and constraints}

\begin{figure}
	\centering
	\includegraphics[scale=1]{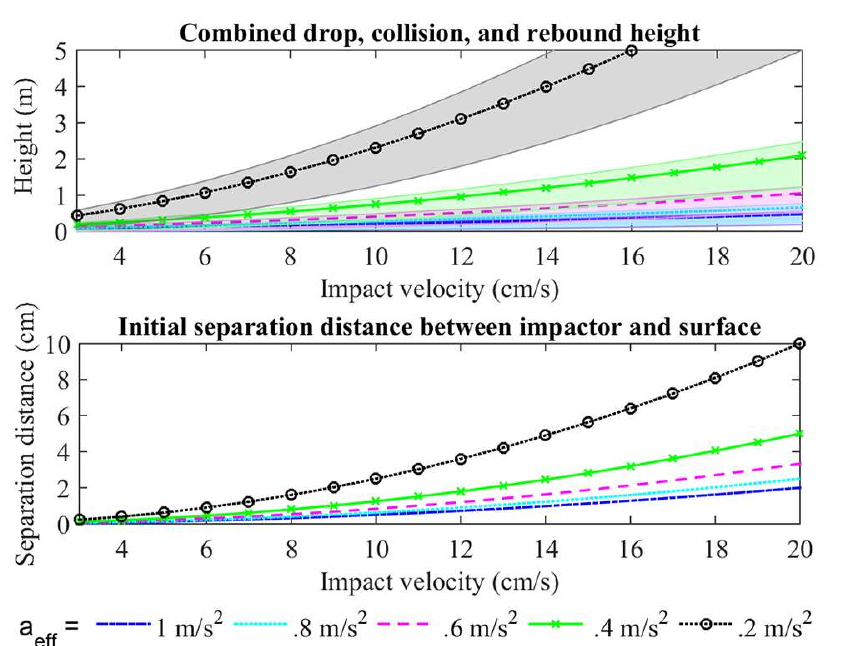}
	\caption[Experimental drop height for different collision velocities and effective accelerations]{Top: experimental height for different effective surface accelerations and collision velocities. Shading indicates the range of possible height values caused by estimating the coefficient of restitution and collision time. The required experiment height increases as the desired effective acceleration decreases. Bottom: initial separation distance between the projectile and surface material for different effective surface accelerations and collision velocities. The initial separation distance decreases as the desired effective acceleration increases.}
	\label{fig:totalHeight}
\end{figure}

Figure \ref{fig:totalHeight} shows how the separation distance between the projectile and surface material and how the required experimental height, from the initial release to the beginning of the deceleration phase, evolves for different effective surface accelerations and collision velocities. The curves on the plot are constructed using Equation \ref{eq:fullHeight}, and the shading indicates the range of possible height values caused by estimating the coefficient of restitution and collision time in the calculations, as discussed in Section \ref{estParameters}. The plot shoes that, for a given impact velocity, the required height increases as the effective acceleration decreases. For a given impact velocity, initial separation distance decreases as the effective acceleration increases.  As a consequence, the lowest effective acceleration that can be observed, for a given collision velocity in the experiment, is constrained by the drop tower's physical height, and the highest effective acceleration that can be observed is limited to a separation distance that is physically reasonable to setup and measure.

\section{Results}
\label{Results}


The performance of the experimental set-up is analyzed based on a number of qualitative and quantitative factors, such as projectile and surface material visibility, surface container deceleration, system friction, and release synchronization. The descent of the surface container is of particular interest, because the guide rail, pulley, and counterweight guide tube features of the experiment cause the desired acceleration of the surface container to deviate from its theoretical value.

\subsection{Experiment performance and repeatability}

\begin{figure}
	\centering
	\includegraphics[scale=1]{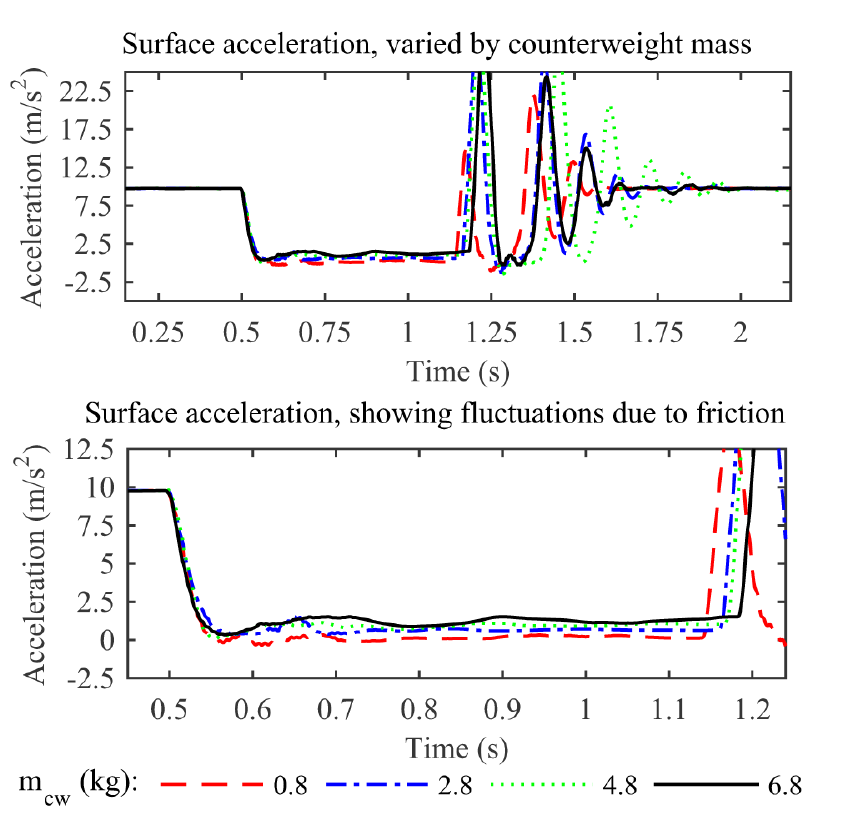}
	\caption[Friction Plot]{Top: acceleration profile for a 160 kg, sand-filled container dropped from a height of 2.20 m. The effective acceleration of the surface material increases as counterweight mass increases. Bottom: zoomed-in view of the surface container's vertical acceleration over the free-fall period. Due to friction, the container's acceleration fluctuates along the length of the rails, though more so at the top of the rails than the bottom.}
	\label{fig:frictionPlot}
\end{figure}

The difference between the surface container's actual and predicted acceleration is analyzed using accelerometer data from drop tests with varied counterweight masses. The top plot in Figure \ref{fig:frictionPlot} shows the container's vertical acceleration profile for a sand-filled container dropped from a height of 2.20 m and resisted by counterweight masses of 0.8, 2.8, 4.8, and 6.8 kg. The data was smoothed using a low pass filter with a cut-off frequency of 10 Hz. As expected, the surface material's effective acceleration decreases with increasing counterweight mass. Based on Equation \ref{eq:cwMass}, a 160 kg surface container should have accelerations of 9.71, 9.47, 9.24, and 9.01 m/s\textsuperscript{2} for counterweight masses of 0.8, 2.8, 4.8, and 6.8 kg respectively. The container's acceleration during the experimental trials is determined by averaging the raw accelerometer data during the container's free-fall period, excluding fluctuations from the container's initial release. Table \ref{tab:counterweightConclusions} lists the theoretical and experimental accelerations of the surface container for the different trials depicted in Figure \ref{fig:frictionPlot}. The percent difference between the actual and predicted accelerations range from 0.93-4.31\%, increasing with increasing counterweight mass.

Also as expected, the container's free-fall duration increases as more counterweights are added to the system. The duration of the free-fall phase can be estimated from the experimental trials by taking the elapsed time between the peaks in the accelerometer data that indicate the container's initial release and then its impact with the honeycomb panels. Table \ref{tab:counterweightConclusions} lists the container's estimated free-fall duration for the 2.0 m drop tests.

\begin{table}
\caption{Comparison of theoretical acceleration ($a_{th}$) and experimental accelerations ($a_{ex}$) for a 160 kg, sand-filled surface container dropped from a height of 2.0 m. The container's experimental acceleration is determined by averaging the raw accelerometer data over the container's free-fall period, excluding fluctuations from the container's initial release. The experimental free-fall duration ($t_{ex}$) is estimated by taking the elapsed time between the peaks in the accelerometer data that indicate the initial release and then impact with the honeycomb panels.}
\label{tab:counterweightConclusions}
\begin{ruledtabular}
\begin{tabular}{cccccccc}
 $m_{cw}$ (kg) & $t_{ex}$ (s) & $a_{th}$ (m/s\textsuperscript{2}) & $a_{ex}$ (m/s\textsuperscript{2}) & \% diff in $a$ \\ \hline
	0.8 & 0.645 & 9.71 & 9.62 & 0.93 \\ 
	2.8 & 0.667 & 9.47 & 9.16 & 3.38 \\ 
	4.8 & 0.674 & 9.24 & 8.87 & 4.02 \\ 
	6.8 & 0.684 & 9.01 & 8.63 & 4.31 \\ 
\end{tabular}
\end{ruledtabular}
\end{table}

The accelerometer data from these trials is also used to make observations about the combined friction in the guide rails, pulleys, and counterweight guide tubes. The bottom plot in Figure \ref{fig:frictionPlot} shows a zoomed-in view of the surface container's Y-axis acceleration over the free-fall period. After the container is released, its acceleration fluctuates along the length of the rails, though more so at the top of the rails than the bottom. In addition to friction and cord elasticity, these fluctuations may be caused by deformation in the rails or back-and-forth tilting of the surface container as it falls.

\begin{figure}
	\centering
	\includegraphics[scale=1.0]{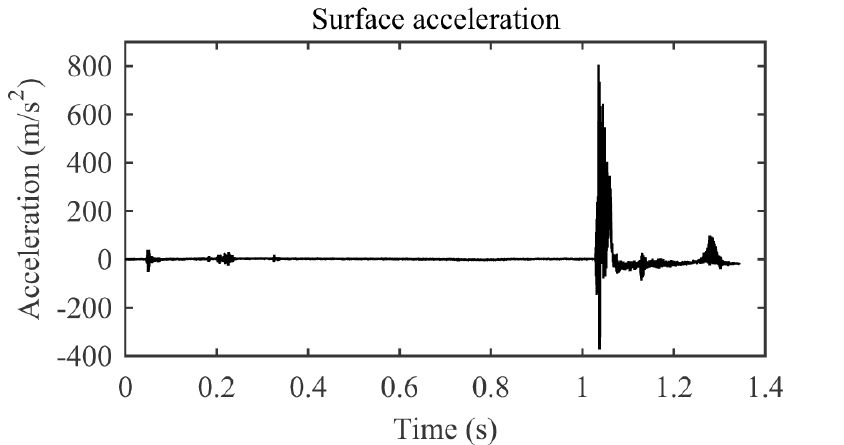}
	\caption[Deceleration Profile]{Deceleration profile of the surface container at the end of a 3 meter drop, with normalized acceleration on the y-axis and time on the x-axis. The surface container experiences a shock of about 80 g upon impacting two 10 cm thick honeycomb panels.}
	\label{fig:decelProfile}
\end{figure}

The surface container's deceleration profile is used to quantify the container's shock upon impact with the honeycomb and to verify that the container is brought to a controlled stop. Four drop tests were performed at a height of 3 meters to verify the functionality of the deceleration system. For these tests, a KISTLER type 8704B500 accelerometer with a range of $\pm$ 500$g$ and sensitivity of 10.4 mV/g was fixed on the surface container in order to record the container's deceleration profile. As seen in Figure \ref{fig:decelProfile}, the surface container experiences a shock of about 80$g$ at impact. The honeycomb panels successfully prevent the container from becoming damaged after the fall. 

Accelerometers installed inside of the surface container and projectile are used to analyze the consistency of the electromagnetic release mechanism. Like friction, the conditions of the projectile's release influence the calibration of the container's starting height. If the release of the projectile is not synchronized with the release of the container, than the delay needs to be accounted for so that the projectile's collision with the surface will take place within the camera's fixed viewing window. Though the release of the surface container and projectile are not perfectly synchronous, initial testing reveals that no additional calibration is required to compensate for the difference.

\begin{figure}
	\centering
	\includegraphics[scale=1]{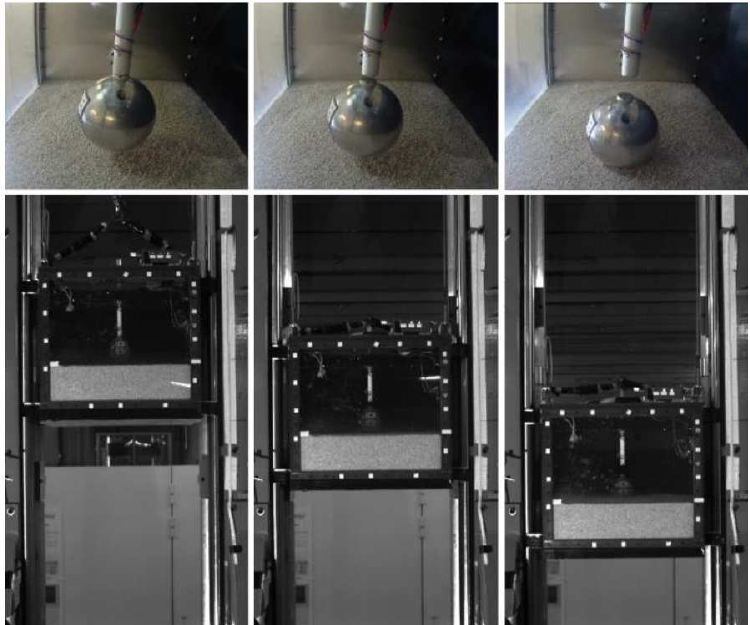}
	\caption[Drop test with high-speed camera]{Images sequence depicting the initial release, free-fall and collision phases for a 2.0 m drop test. The initial separation distance between the projectile and surface material is $\sim$ 3.0 cm. The top row of images are from an AEE MagiCam SD100 camera that was mounted inside of the surface container, and the bottom row of images are still frames from the Ultima APX-RS Photron high-speed camera.}
	\label{fig:cameraStills}
\end{figure}

During several of the performance tests, the surface container was partially filled with sand, and high speed cameras were used to capture the container's descent. The images are used to verify that the sand does not lift and obstruct the view of the projectile during the fall. Figure \ref{fig:cameraStills} shows a sequence of images for a 2.0 m drop test. For this test, the initial separation distance between the projectile and surface material is $\sim$ 3.0 cm. The projectile is clearly visible within the camera frame throughout the release, free-fall, and collision phases of the trial.

These trials are also used to study the packing of the surface material upon impact with the honeycomb. If the sand level changes between the start and finish of the experiment, then a method to re-prepare the sand is required so that the bulk density of the surface material remains consistent between trials. In the calibration tests, it was observed that the sand was no longer level at the end of the deceleration phase. Once the sand was brushed however, its level measurement was unchanged from its value at the beginning of the trial. One hypothesis is that the sand compresses during its initial shock, but becomes disturbed and uncompressed during the oscillations that proceed the shock. The deceleration system naturally regulates the bulk density of the surface material, indicating that no extra treatment of the surface sample is likely to be required between trials.

\section{Discussion}
\label{Discussion}

\begin{figure}
	\centering
	\includegraphics[scale=1]{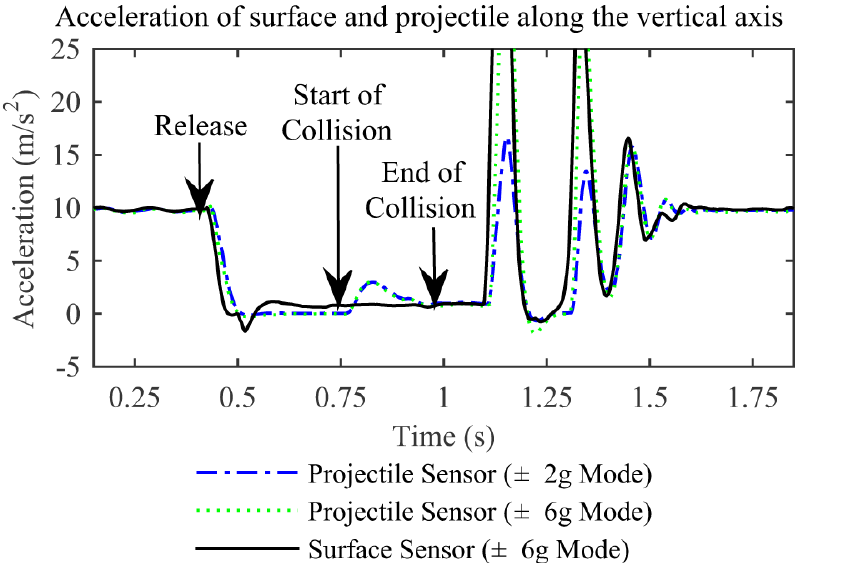}
	\caption[Example Trial]{Acceleration profile of the projectile and surface container for a scientific trial with a starting drop height of 2.20 m, an initial separation distance between the projectile and surface of 3.0 cm, and a counterweight mass of 4.8 kg. The surface has an effective acceleration of approximately 0.82 m/s\textsuperscript{2}. The collision phase lasts for approximately 0.19 seconds and does not result in a detectable rebound.}
	\label{fig:ExampleTrial}
\end{figure}

Preliminary verification tests indicate that the experiment setup accomplishes its primary purpose: to provide a surface container with a controllable downward acceleration so that a free-falling projectile may impact a surface in a reduced-gravity state. Figure \ref{fig:ExampleTrial} shows the acceleration profiles of the projectile and surface container for a scientific trial with a starting drop height of 2.20 m, an initial separation distance between the projectile and surface of 3 cm, and a counterweight mass of 4.8 kg. The data was smoothed using a low pass filter in MATLAB, with a cut-off frequency of 10 Hz. In this trial, the surface has an effective acceleration of approximately 0.82 m/s\textsuperscript{2}. The surface container and projectile are released from rest around the 0.40 second mark. Then, the projectile and surface container are in free-fall until the projectile collides with the surface at 0.75 seconds. The collision phase lasts for approximately 0.19 seconds, or nearly twice as long as the duration used in theoretical calculations, and does not result in a detectable rebound. After an estimated 0.67 seconds of free-fall, the surface container collides with the honeycomb panels and begins to decelerate. A complete analysis of the collision dynamics from this trial, as well as other trials with varying parameters, will be part of a future study.

The combined realities of friction and air drag cause the container's downward acceleration to deviate from its theoretical value. There is also a significant shock that can observed on the surface sensor (at $\sim$0.5 s in Figure \ref{fig:ExampleTrial}) shortly after the release when the chain, visible in Figure \ref{fig:container}, falls onto the box. This is, however, not problematic as long as the chain impact occurs well before the projectile collides with the sand. In the experiment's current state, friction is difficult to correct for because it varies significantly with time. However, to reduce the impacts of friction on the container's acceleration profile, a lubricate can be added and re-applied to the guide rails between trials. The experimental trials can also be configured so that the drop height is skewed towards the bottom of the tower, where the container's acceleration profile tends to fluctuate the least.

Without adding additional counterweight mass to the system, the surface container is naturally slowed to a downward acceleration of approximately 0.1 m/s\textsuperscript{2}, marking the lowest possible effective acceleration that the set-up can achieve. This acceleration is within the scope of planned trials and is less than that of asteroids (1) Ceres and (4) Vesta, with surface gravities of ~0.29 m/s\textsuperscript{2} and ~0.25 m/s\textsuperscript{2}, respectively \cite{Russell2012, Carry2008}. The largest effective acceleration that has currently been tested, 1 m/s\textsuperscript{2} is comparable to the surface gravity of Saturn's moon Enceladus. 

Accelerations lower than 0.1 m/s\textsuperscript{2} can be achieved by re-designing the set-up for a free-falling container. This would eliminate the constraints that rail friction imposes on the system. In addition, the unique experimental data obtained in these trials may also be valuable to benchmark different numerical simulation approaches (Distinct Element Method, Finite Element Method, etc.). These simulations can then subsequently be used to extrapolate the results to even lower gravity regimes.

Measurements of the surface material's level before and after each trial indicate that the system inherently disturbs the material during the deceleration process, effectively holding the bulk density of the surface material constant between trials. If as few as 2 trials are performed on a given day, then no additional actions are required to prepare the sand between each trial. However, the level of the sand should still be monitored between trials so that it can be treated in the case that packing is observed.  

In addition to providing quantitative insight into the setup's performance, calibration trials help configure the setup's data acquisition tools and ground support equipment. These activities include optimization of camera placement, camera lighting, and camera and accelerometer settings.

\section{Conclusions}
\label{Conclusions}


The presented design has been used to study low-velocity collisions into granular surfaces. The first round of testing used a 10 cm metallic sphere as a projectile and sand as a surface simulant. The collision velocities and effective accelerations that can be attained in this setup are constrained by the size and physical limitations of the drop-tower structure. The system can currently support effective accelerations of $\sim$0.1 to 1 m/s\textsuperscript{2}, though higher values can be reached by simply adding more counterweight mass to the pulley sub-system. Other test parameters are easily adjustable and open-up possibilities for future experimentation. For example, the release mechanism can support changes to the size, mass, and shape of the projectile, while the container can hold a range of surface simulant depths and materials. 

Since the surface materials in this set-up are subject to the effects of gravity at all times, this set-up does not have the same concern for particulate lift as parabolic flight experiments. Therefore, there are no limitations from the perspective of the initial release as to the size or material of granular matter that can be used. However, it should be noted that air resistance may be a limiting factor in grain-size selection. The interstitial air effect is negligible for grains with diameters $>$ 0.1 mm \cite{Katsuragi2016, Pak1995}. This is demonstrated specifically for low-speed impacts into a granular material by Katsuragi and Durian (2007) \cite{katsuragi2007}. The only constraint to the type of solid material that can be used in the current set-up is that it must survive the shock of deceleration.

Outside of collision experiments, this setup can be used to test the general strength and structural properties of granular materials in reduced-gravity environments. For instance, the surface container can be redesigned to incorporate penetrometers and shear tools. The container can also be redesigned to accept entire mobility or sampling mechanisms in order to validate the mechanism's at-rest and in-motion surface interactions. For these tests, the particle size and effective acceleration of the surface material can be altered to mimic the properties of small bodies or large asteroids and comets. The limiting factor in using this setup for other experiments is the short duration of the drop time. If the surface container free-falls (i.e., no counterweights) from the maximum possible drop height of 5 meters, only 1 second elapses before the container impacts the deceleration material. 

The drop tower structure, the counterweight and pulley system, and the deceleration system are stand-alone components of the experiment in the sense that they do not require any redesign or re-calibration in order to be used for other tests. This aspect of the experiment makes it ideal for frequent, inexpensive, and repeatable experimentation.

\begin{acknowledgments}

This project benefited from some financial support from the Centre National d'Etudes Spatiales (CNES) and was a collaborative effort between several departments at ISAE-SUPAERO. Alexandre Cadu and Anthony Sournac (both DEOS/SSPA) provided extremely valuable help and advice concerning the installation of the electromagnetic release system. Emmanuel Zenou, from the D\'{e}partement d'Ing\'{e}nierie des Syst\`{e}mes Complexes (DISC), helped prepare the experiment in such a way to facilitate the image analyses during the data processing. Lastly, Daniel Gagneux and Thierry Faure, from the D\'{e}partement M\'{e}canique des Structures et Mat\'{e}riaux (DMSM), completed the experiment's detailed design and fabrication.

\end{acknowledgments}

\bibliography{bibFile}

\begin{thebibliography}{45}%
\makeatletter
\providecommand \@ifxundefined [1]{%
 \@ifx{#1\undefined}
}%
\providecommand \@ifnum [1]{%
 \ifnum #1\expandafter \@firstoftwo
 \else \expandafter \@secondoftwo
 \fi
}%
\providecommand \@ifx [1]{%
 \ifx #1\expandafter \@firstoftwo
 \else \expandafter \@secondoftwo
 \fi
}%
\providecommand \natexlab [1]{#1}%
\providecommand \enquote  [1]{``#1''}%
\providecommand \bibnamefont  [1]{#1}%
\providecommand \bibfnamefont [1]{#1}%
\providecommand \citenamefont [1]{#1}%
\providecommand \href@noop [0]{\@secondoftwo}%
\providecommand \href [0]{\begingroup \@sanitize@url \@href}%
\providecommand \@href[1]{\@@startlink{#1}\@@href}%
\providecommand \@@href[1]{\endgroup#1\@@endlink}%
\providecommand \@sanitize@url [0]{\catcode `\\12\catcode `\$12\catcode
  `\&12\catcode `\#12\catcode `\^12\catcode `\_12\catcode `\%12\relax}%
\providecommand \@@startlink[1]{}%
\providecommand \@@endlink[0]{}%
\providecommand \url  [0]{\begingroup\@sanitize@url \@url }%
\providecommand \@url [1]{\endgroup\@href {#1}{\urlprefix }}%
\providecommand \urlprefix  [0]{URL }%
\providecommand \Eprint [0]{\href }%
\providecommand \doibase [0]{http://dx.doi.org/}%
\providecommand \selectlanguage [0]{\@gobble}%
\providecommand \bibinfo  [0]{\@secondoftwo}%
\providecommand \bibfield  [0]{\@secondoftwo}%
\providecommand \translation [1]{[#1]}%
\providecommand \BibitemOpen [0]{}%
\providecommand \bibitemStop [0]{}%
\providecommand \bibitemNoStop [0]{.\EOS\space}%
\providecommand \EOS [0]{\spacefactor3000\relax}%
\providecommand \BibitemShut  [1]{\csname bibitem#1\endcsname}%
\let\auto@bib@innerbib\@empty
\bibitem [{\citenamefont {Biele}\ \emph {et~al.}(2015)\citenamefont {Biele},
  \citenamefont {Ulamec}, \citenamefont {Maibaum}, \citenamefont {Roll},
  \citenamefont {Witte}, \citenamefont {Jurado}, \citenamefont {Mu{\~n}oz},
  \citenamefont {Arnold}, \citenamefont {Auster}, \citenamefont {Casas} \emph
  {et~al.}}]{biele2015landing}%
  \BibitemOpen
  \bibfield  {author} {\bibinfo {author} {\bibfnamefont {J.}~\bibnamefont
  {Biele}}, \bibinfo {author} {\bibfnamefont {S.}~\bibnamefont {Ulamec}},
  \bibinfo {author} {\bibfnamefont {M.}~\bibnamefont {Maibaum}}, \bibinfo
  {author} {\bibfnamefont {R.}~\bibnamefont {Roll}}, \bibinfo {author}
  {\bibfnamefont {L.}~\bibnamefont {Witte}}, \bibinfo {author} {\bibfnamefont
  {E.}~\bibnamefont {Jurado}}, \bibinfo {author} {\bibfnamefont
  {P.}~\bibnamefont {Mu{\~n}oz}}, \bibinfo {author} {\bibfnamefont
  {W.}~\bibnamefont {Arnold}}, \bibinfo {author} {\bibfnamefont {H.-U.}\
  \bibnamefont {Auster}}, \bibinfo {author} {\bibfnamefont {C.}~\bibnamefont
  {Casas}},  \emph {et~al.},\ }\href@noop {} {\bibfield  {journal} {\bibinfo
  {journal} {Science}\ }\textbf {\bibinfo {volume} {349}},\ \bibinfo {pages}
  {aaa9816} (\bibinfo {year} {2015})}\BibitemShut {NoStop}%
\bibitem [{\citenamefont {Ulamec}\ \emph {et~al.}()\citenamefont {Ulamec},
  \citenamefont {Fantinati}, \citenamefont {Maibaum}, \citenamefont {Geurts},
  \citenamefont {Biele}, \citenamefont {Jansen}, \citenamefont {K\"{u}chemann},
  \citenamefont {Cozzoni}, \citenamefont {Finke}, \citenamefont {Lommatsch},
  \citenamefont {Moussi-Soffys}, \citenamefont {Delmas},\ and\ \citenamefont
  {O'Rourke}}]{Ulamec2015}%
  \BibitemOpen
  \bibfield  {author} {\bibinfo {author} {\bibfnamefont {S.}~\bibnamefont
  {Ulamec}}, \bibinfo {author} {\bibfnamefont {C.}~\bibnamefont {Fantinati}},
  \bibinfo {author} {\bibfnamefont {M.}~\bibnamefont {Maibaum}}, \bibinfo
  {author} {\bibfnamefont {K.}~\bibnamefont {Geurts}}, \bibinfo {author}
  {\bibfnamefont {J.}~\bibnamefont {Biele}}, \bibinfo {author} {\bibfnamefont
  {S.}~\bibnamefont {Jansen}}, \bibinfo {author} {\bibfnamefont
  {O.}~\bibnamefont {K\"{u}chemann}}, \bibinfo {author} {\bibfnamefont
  {B.}~\bibnamefont {Cozzoni}}, \bibinfo {author} {\bibfnamefont
  {F.}~\bibnamefont {Finke}}, \bibinfo {author} {\bibfnamefont
  {V.}~\bibnamefont {Lommatsch}}, \bibinfo {author} {\bibfnamefont
  {A.}~\bibnamefont {Moussi-Soffys}}, \bibinfo {author} {\bibfnamefont
  {C.}~\bibnamefont {Delmas}}, \ and\ \bibinfo {author} {\bibfnamefont
  {L.}~\bibnamefont {O'Rourke}},\ }\href@noop {} {\bibinfo  {journal} {Acta
  Astronautica}\ }\BibitemShut {NoStop}%
\bibitem [{\citenamefont {Tsuda}\ \emph {et~al.}(2013)\citenamefont {Tsuda},
  \citenamefont {Yoshikawa}, \citenamefont {Abe}, \citenamefont {Minamino},\
  and\ \citenamefont {Nakazawa}}]{Tsuda}%
  \BibitemOpen
\bibfield  {journal} {  }\bibfield  {author} {\bibinfo {author} {\bibfnamefont
  {Y.}~\bibnamefont {Tsuda}}, \bibinfo {author} {\bibfnamefont
  {M.}~\bibnamefont {Yoshikawa}}, \bibinfo {author} {\bibfnamefont
  {M.}~\bibnamefont {Abe}}, \bibinfo {author} {\bibfnamefont {H.}~\bibnamefont
  {Minamino}}, \ and\ \bibinfo {author} {\bibfnamefont {S.}~\bibnamefont
  {Nakazawa}},\ }\href@noop {} {\bibfield  {journal} {\bibinfo  {journal} {Acta
  Astronautica}\ }\textbf {\bibinfo {volume} {90}},\ \bibinfo {pages} {356 }
  (\bibinfo {year} {2013})}\BibitemShut {NoStop}%
\bibitem [{\citenamefont {Ho}\ \emph {et~al.}(2014)\citenamefont {Ho},
  \citenamefont {Findlay}, \citenamefont {Ziach}, \citenamefont {Krause},
  \citenamefont {Lange}, \citenamefont {Reill}, \citenamefont {Deleuze},
  \citenamefont {Ulamec}, \citenamefont {Biele}, \citenamefont {Jaumann} \emph
  {et~al.}}]{Ho}%
  \BibitemOpen
  \bibfield  {author} {\bibinfo {author} {\bibfnamefont {T.}~\bibnamefont
  {Ho}}, \bibinfo {author} {\bibfnamefont {R.}~\bibnamefont {Findlay}},
  \bibinfo {author} {\bibfnamefont {C.}~\bibnamefont {Ziach}}, \bibinfo
  {author} {\bibfnamefont {C.}~\bibnamefont {Krause}}, \bibinfo {author}
  {\bibfnamefont {M.}~\bibnamefont {Lange}}, \bibinfo {author} {\bibfnamefont
  {J.}~\bibnamefont {Reill}}, \bibinfo {author} {\bibfnamefont
  {M.}~\bibnamefont {Deleuze}}, \bibinfo {author} {\bibfnamefont
  {S.}~\bibnamefont {Ulamec}}, \bibinfo {author} {\bibfnamefont
  {J.}~\bibnamefont {Biele}}, \bibinfo {author} {\bibfnamefont
  {R.}~\bibnamefont {Jaumann}},  \emph {et~al.},\ }\href@noop {} {\bibfield
  {journal} {\bibinfo  {journal} {Lunar and Planetary Institute Science
  Conference Abstracts}\ }\textbf {\bibinfo {volume} {45}} (\bibinfo {year}
  {2014})}\BibitemShut {NoStop}%
\bibitem [{\citenamefont {Michel}\ \emph {et~al.}(2015)\citenamefont {Michel},
  \citenamefont {Cheng}, \citenamefont {Ulamec},\ and\ \citenamefont {the
  AIDA~team 2015}}]{michelasteroid}%
  \BibitemOpen
  \bibfield  {author} {\bibinfo {author} {\bibfnamefont {P.}~\bibnamefont
  {Michel}}, \bibinfo {author} {\bibfnamefont {A.}~\bibnamefont {Cheng}},
  \bibinfo {author} {\bibfnamefont {S.}~\bibnamefont {Ulamec}}, \ and\ \bibinfo
  {author} {\bibnamefont {the AIDA~team 2015}}\ }(\bibinfo  {publisher}
  {Planetary Defense Conference},\ \bibinfo {address} {Frascati, Italy},\
  \bibinfo {year} {2015})\BibitemShut {NoStop}%
\bibitem [{\citenamefont {Cheng}\ \emph {et~al.}(2015)\citenamefont {Cheng},
  \citenamefont {Atchison}, \citenamefont {Kantsiper}, \citenamefont {Rivkin},
  \citenamefont {Stickle}, \citenamefont {Reed}, \citenamefont {Galvez},
  \citenamefont {Carnelli}, \citenamefont {Michel},\ and\ \citenamefont
  {Ulamec}}]{cheng2015asteroid}%
  \BibitemOpen
  \bibfield  {author} {\bibinfo {author} {\bibfnamefont {A.~F.}\ \bibnamefont
  {Cheng}}, \bibinfo {author} {\bibfnamefont {J.}~\bibnamefont {Atchison}},
  \bibinfo {author} {\bibfnamefont {B.}~\bibnamefont {Kantsiper}}, \bibinfo
  {author} {\bibfnamefont {A.~S.}\ \bibnamefont {Rivkin}}, \bibinfo {author}
  {\bibfnamefont {A.}~\bibnamefont {Stickle}}, \bibinfo {author} {\bibfnamefont
  {C.}~\bibnamefont {Reed}}, \bibinfo {author} {\bibfnamefont {A.}~\bibnamefont
  {Galvez}}, \bibinfo {author} {\bibfnamefont {I.}~\bibnamefont {Carnelli}},
  \bibinfo {author} {\bibfnamefont {P.}~\bibnamefont {Michel}}, \ and\ \bibinfo
  {author} {\bibfnamefont {S.}~\bibnamefont {Ulamec}},\ }\href@noop {}
  {\bibfield  {journal} {\bibinfo  {journal} {Acta Astronautica}\ }\textbf
  {\bibinfo {volume} {115}},\ \bibinfo {pages} {262} (\bibinfo {year}
  {2015})}\BibitemShut {NoStop}%
\bibitem [{\citenamefont {Harris}(2005)}]{harris2005surface}%
  \BibitemOpen
  \bibfield  {author} {\bibinfo {author} {\bibfnamefont {A.~W.}\ \bibnamefont
  {Harris}},\ }\href@noop {} {\bibfield  {journal} {\bibinfo  {journal}
  {Proceedings of the International Astronomical Union}\ }\textbf {\bibinfo
  {volume} {1}},\ \bibinfo {pages} {449} (\bibinfo {year} {2005})}\BibitemShut
  {NoStop}%
\bibitem [{\citenamefont {Campins}\ \emph {et~al.}(2009)\citenamefont
  {Campins}, \citenamefont {Emery}, \citenamefont {Kelley}, \citenamefont
  {Fern{\'a}ndez}, \citenamefont {Licandro}, \citenamefont {Delb{\'o}},
  \citenamefont {Barucci},\ and\ \citenamefont {Dotto}}]{Campins}%
  \BibitemOpen
  \bibfield  {author} {\bibinfo {author} {\bibfnamefont {H.}~\bibnamefont
  {Campins}}, \bibinfo {author} {\bibfnamefont {J.~P.}\ \bibnamefont {Emery}},
  \bibinfo {author} {\bibfnamefont {M.}~\bibnamefont {Kelley}}, \bibinfo
  {author} {\bibfnamefont {Y.}~\bibnamefont {Fern{\'a}ndez}}, \bibinfo {author}
  {\bibfnamefont {J.}~\bibnamefont {Licandro}}, \bibinfo {author}
  {\bibfnamefont {M.}~\bibnamefont {Delb{\'o}}}, \bibinfo {author}
  {\bibfnamefont {A.}~\bibnamefont {Barucci}}, \ and\ \bibinfo {author}
  {\bibfnamefont {E.}~\bibnamefont {Dotto}},\ }\href@noop {} {\bibfield
  {journal} {\bibinfo  {journal} {Astronomy and Astrophysics}\ } (\bibinfo
  {year} {2009})}\BibitemShut {NoStop}%
\bibitem [{\citenamefont {Gundlach}\ and\ \citenamefont
  {Blum}(2013)}]{Gundlach}%
  \BibitemOpen
  \bibfield  {author} {\bibinfo {author} {\bibfnamefont {B.}~\bibnamefont
  {Gundlach}}\ and\ \bibinfo {author} {\bibfnamefont {J.}~\bibnamefont
  {Blum}},\ }\href@noop {} {\bibfield  {journal} {\bibinfo  {journal} {Icarus}\
  }\textbf {\bibinfo {volume} {223}},\ \bibinfo {pages} {479} (\bibinfo {year}
  {2013})}\BibitemShut {NoStop}%
\bibitem [{\citenamefont {Moskovitz}\ \emph {et~al.}(2013)\citenamefont
  {Moskovitz}, \citenamefont {Abe}, \citenamefont {Pan}, \citenamefont {Osip},
  \citenamefont {Pefkou}, \citenamefont {Melita}, \citenamefont {Elias},
  \citenamefont {Kitazato}, \citenamefont {Bus}, \citenamefont {DeMeo} \emph
  {et~al.}}]{Moskovitz}%
  \BibitemOpen
  \bibfield  {author} {\bibinfo {author} {\bibfnamefont {N.~A.}\ \bibnamefont
  {Moskovitz}}, \bibinfo {author} {\bibfnamefont {S.}~\bibnamefont {Abe}},
  \bibinfo {author} {\bibfnamefont {K.-S.}\ \bibnamefont {Pan}}, \bibinfo
  {author} {\bibfnamefont {D.~J.}\ \bibnamefont {Osip}}, \bibinfo {author}
  {\bibfnamefont {D.}~\bibnamefont {Pefkou}}, \bibinfo {author} {\bibfnamefont
  {M.~D.}\ \bibnamefont {Melita}}, \bibinfo {author} {\bibfnamefont
  {M.}~\bibnamefont {Elias}}, \bibinfo {author} {\bibfnamefont
  {K.}~\bibnamefont {Kitazato}}, \bibinfo {author} {\bibfnamefont {S.~J.}\
  \bibnamefont {Bus}}, \bibinfo {author} {\bibfnamefont {F.~E.}\ \bibnamefont
  {DeMeo}},  \emph {et~al.},\ }\href@noop {} {\bibfield  {journal} {\bibinfo
  {journal} {Icarus}\ }\textbf {\bibinfo {volume} {224}},\ \bibinfo {pages}
  {24} (\bibinfo {year} {2013})}\BibitemShut {NoStop}%
\bibitem [{\citenamefont {Sullivan}\ \emph {et~al.}(1996)\citenamefont
  {Sullivan}, \citenamefont {Greeley}, \citenamefont {Pappalardo},
  \citenamefont {Asphaug}, \citenamefont {Moore}, \citenamefont {Morrison},
  \citenamefont {Belton}, \citenamefont {Carr}, \citenamefont {Chapman},
  \citenamefont {Geissler} \emph {et~al.}}]{sullivan1996geology}%
  \BibitemOpen
  \bibfield  {author} {\bibinfo {author} {\bibfnamefont {R.}~\bibnamefont
  {Sullivan}}, \bibinfo {author} {\bibfnamefont {R.}~\bibnamefont {Greeley}},
  \bibinfo {author} {\bibfnamefont {R.}~\bibnamefont {Pappalardo}}, \bibinfo
  {author} {\bibfnamefont {E.}~\bibnamefont {Asphaug}}, \bibinfo {author}
  {\bibfnamefont {J.~M.}\ \bibnamefont {Moore}}, \bibinfo {author}
  {\bibfnamefont {D.}~\bibnamefont {Morrison}}, \bibinfo {author}
  {\bibfnamefont {M.~J.~S.}\ \bibnamefont {Belton}}, \bibinfo {author}
  {\bibfnamefont {M.}~\bibnamefont {Carr}}, \bibinfo {author} {\bibfnamefont
  {C.~R.}\ \bibnamefont {Chapman}}, \bibinfo {author} {\bibfnamefont
  {P.}~\bibnamefont {Geissler}},  \emph {et~al.},\ }\href@noop {} {\bibfield
  {journal} {\bibinfo  {journal} {Icarus}\ }\textbf {\bibinfo {volume} {120}},\
  \bibinfo {pages} {119} (\bibinfo {year} {1996})}\BibitemShut {NoStop}%
\bibitem [{\citenamefont {Carr}\ \emph {et~al.}(1994)\citenamefont {Carr},
  \citenamefont {Kirk}, \citenamefont {McEwen}, \citenamefont {Veverka},
  \citenamefont {Thomas}, \citenamefont {Head},\ and\ \citenamefont
  {Murchie}}]{carr1994geology}%
  \BibitemOpen
  \bibfield  {author} {\bibinfo {author} {\bibfnamefont {M.~H.}\ \bibnamefont
  {Carr}}, \bibinfo {author} {\bibfnamefont {R.~L.}\ \bibnamefont {Kirk}},
  \bibinfo {author} {\bibfnamefont {A.}~\bibnamefont {McEwen}}, \bibinfo
  {author} {\bibfnamefont {J.}~\bibnamefont {Veverka}}, \bibinfo {author}
  {\bibfnamefont {P.~H. J.~W.}\ \bibnamefont {Thomas}}, \bibinfo {author}
  {\bibfnamefont {J.~W.}\ \bibnamefont {Head}}, \ and\ \bibinfo {author}
  {\bibfnamefont {S.}~\bibnamefont {Murchie}},\ }\href@noop {} {\bibfield
  {journal} {\bibinfo  {journal} {Icarus}\ }\textbf {\bibinfo {volume} {107}},\
  \bibinfo {pages} {61} (\bibinfo {year} {1994})}\BibitemShut {NoStop}%
\bibitem [{\citenamefont {Thomas}\ \emph {et~al.}(1999)\citenamefont {Thomas},
  \citenamefont {Veverka}, \citenamefont {Bell}, \citenamefont {Clark},
  \citenamefont {Carcich}, \citenamefont {Joseph}, \citenamefont {Robinson},
  \citenamefont {McFadden}, \citenamefont {Malin}, \citenamefont {Chapman}
  \emph {et~al.}}]{thomas1999mathilde}%
  \BibitemOpen
  \bibfield  {author} {\bibinfo {author} {\bibfnamefont {P.~C.}\ \bibnamefont
  {Thomas}}, \bibinfo {author} {\bibfnamefont {J.}~\bibnamefont {Veverka}},
  \bibinfo {author} {\bibfnamefont {J.~F.}\ \bibnamefont {Bell}}, \bibinfo
  {author} {\bibfnamefont {B.~E.}\ \bibnamefont {Clark}}, \bibinfo {author}
  {\bibfnamefont {B.}~\bibnamefont {Carcich}}, \bibinfo {author} {\bibfnamefont
  {J.}~\bibnamefont {Joseph}}, \bibinfo {author} {\bibfnamefont
  {M.}~\bibnamefont {Robinson}}, \bibinfo {author} {\bibfnamefont {L.~A.}\
  \bibnamefont {McFadden}}, \bibinfo {author} {\bibfnamefont {M.~C.}\
  \bibnamefont {Malin}}, \bibinfo {author} {\bibfnamefont {C.~R.}\ \bibnamefont
  {Chapman}},  \emph {et~al.},\ }\href@noop {} {\bibfield  {journal} {\bibinfo
  {journal} {Icarus}\ }\textbf {\bibinfo {volume} {140}},\ \bibinfo {pages}
  {17} (\bibinfo {year} {1999})}\BibitemShut {NoStop}%
\bibitem [{\citenamefont {Robinson}\ \emph {et~al.}(2002)\citenamefont
  {Robinson}, \citenamefont {Thomas}, \citenamefont {Veverka}, \citenamefont
  {Murchie},\ and\ \citenamefont {Wilcox}}]{robinson2002geology}%
  \BibitemOpen
  \bibfield  {author} {\bibinfo {author} {\bibfnamefont {M.~S.}\ \bibnamefont
  {Robinson}}, \bibinfo {author} {\bibfnamefont {P.~C.}\ \bibnamefont
  {Thomas}}, \bibinfo {author} {\bibfnamefont {J.}~\bibnamefont {Veverka}},
  \bibinfo {author} {\bibfnamefont {S.~L.}\ \bibnamefont {Murchie}}, \ and\
  \bibinfo {author} {\bibfnamefont {B.~B.}\ \bibnamefont {Wilcox}},\
  }\href@noop {} {\bibfield  {journal} {\bibinfo  {journal} {Meteoritics and
  Planetary Science}\ }\textbf {\bibinfo {volume} {37}},\ \bibinfo {pages}
  {1651} (\bibinfo {year} {2002})}\BibitemShut {NoStop}%
\bibitem [{\citenamefont {Veverka}\ \emph {et~al.}(2000)\citenamefont
  {Veverka}, \citenamefont {Robinson}, \citenamefont {Thomas}, \citenamefont
  {Murchie}, \citenamefont {Bell}, \citenamefont {Izenberg}, \citenamefont
  {Chapman}, \citenamefont {Harch}, \citenamefont {Bell}, \citenamefont
  {Carcich} \emph {et~al.}}]{Veverka}%
  \BibitemOpen
  \bibfield  {author} {\bibinfo {author} {\bibfnamefont {J.}~\bibnamefont
  {Veverka}}, \bibinfo {author} {\bibfnamefont {M.}~\bibnamefont {Robinson}},
  \bibinfo {author} {\bibfnamefont {P.}~\bibnamefont {Thomas}}, \bibinfo
  {author} {\bibfnamefont {S.}~\bibnamefont {Murchie}}, \bibinfo {author}
  {\bibfnamefont {J.~F.}\ \bibnamefont {Bell}}, \bibinfo {author}
  {\bibfnamefont {N.}~\bibnamefont {Izenberg}}, \bibinfo {author}
  {\bibfnamefont {C.}~\bibnamefont {Chapman}}, \bibinfo {author} {\bibfnamefont
  {A.}~\bibnamefont {Harch}}, \bibinfo {author} {\bibfnamefont
  {M.}~\bibnamefont {Bell}}, \bibinfo {author} {\bibfnamefont {B.}~\bibnamefont
  {Carcich}},  \emph {et~al.},\ }\href@noop {} {\bibfield  {journal} {\bibinfo
  {journal} {Science}\ }\textbf {\bibinfo {volume} {289}},\ \bibinfo {pages}
  {2088} (\bibinfo {year} {2000})}\BibitemShut {NoStop}%
\bibitem [{\citenamefont {Yano}\ \emph {et~al.}(2006)\citenamefont {Yano},
  \citenamefont {Kubota}, \citenamefont {Miyamoto}, \citenamefont {Okada},
  \citenamefont {Scheeres}, \citenamefont {Takagi}, \citenamefont {Yoshida},
  \citenamefont {Abe}, \citenamefont {Abe}, \citenamefont {Barnouin-Jha} \emph
  {et~al.}}]{Yano}%
  \BibitemOpen
  \bibfield  {author} {\bibinfo {author} {\bibfnamefont {H.}~\bibnamefont
  {Yano}}, \bibinfo {author} {\bibfnamefont {T.}~\bibnamefont {Kubota}},
  \bibinfo {author} {\bibfnamefont {H.}~\bibnamefont {Miyamoto}}, \bibinfo
  {author} {\bibfnamefont {T.}~\bibnamefont {Okada}}, \bibinfo {author}
  {\bibfnamefont {D.}~\bibnamefont {Scheeres}}, \bibinfo {author}
  {\bibfnamefont {Y.}~\bibnamefont {Takagi}}, \bibinfo {author} {\bibfnamefont
  {K.}~\bibnamefont {Yoshida}}, \bibinfo {author} {\bibfnamefont
  {M.}~\bibnamefont {Abe}}, \bibinfo {author} {\bibfnamefont {S.}~\bibnamefont
  {Abe}}, \bibinfo {author} {\bibfnamefont {O.}~\bibnamefont {Barnouin-Jha}},
  \emph {et~al.},\ }\href@noop {} {\bibfield  {journal} {\bibinfo  {journal}
  {Science}\ }\textbf {\bibinfo {volume} {312}},\ \bibinfo {pages} {1350}
  (\bibinfo {year} {2006})}\BibitemShut {NoStop}%
\bibitem [{\citenamefont {Jaumann}\ \emph {et~al.}(2012)\citenamefont
  {Jaumann}, \citenamefont {Williams}, \citenamefont {Buczkowski},
  \citenamefont {Yingst}, \citenamefont {Preusker}, \citenamefont {Hiesinger},
  \citenamefont {Schmedemann}, \citenamefont {Kneissl}, \citenamefont
  {Vincent}, \citenamefont {Blewett} \emph {et~al.}}]{jaumann2012vesta}%
  \BibitemOpen
  \bibfield  {author} {\bibinfo {author} {\bibfnamefont {R.}~\bibnamefont
  {Jaumann}}, \bibinfo {author} {\bibfnamefont {D.~A.}\ \bibnamefont
  {Williams}}, \bibinfo {author} {\bibfnamefont {D.~L.}\ \bibnamefont
  {Buczkowski}}, \bibinfo {author} {\bibfnamefont {R.~A.}\ \bibnamefont
  {Yingst}}, \bibinfo {author} {\bibfnamefont {F.}~\bibnamefont {Preusker}},
  \bibinfo {author} {\bibfnamefont {H.}~\bibnamefont {Hiesinger}}, \bibinfo
  {author} {\bibfnamefont {N.}~\bibnamefont {Schmedemann}}, \bibinfo {author}
  {\bibfnamefont {T.}~\bibnamefont {Kneissl}}, \bibinfo {author} {\bibfnamefont
  {J.~B.}\ \bibnamefont {Vincent}}, \bibinfo {author} {\bibfnamefont {D.~T.}\
  \bibnamefont {Blewett}},  \emph {et~al.},\ }\href@noop {} {\bibfield
  {journal} {\bibinfo  {journal} {Science}\ }\textbf {\bibinfo {volume}
  {336}},\ \bibinfo {pages} {687} (\bibinfo {year} {2012})}\BibitemShut
  {NoStop}%
\bibitem [{\citenamefont {Murdoch}\ \emph {et~al.}(2015)\citenamefont
  {Murdoch}, \citenamefont {Sanchez}, \citenamefont {Schwartz},\ and\
  \citenamefont {Miyamoto}}]{murdoch2015}%
  \BibitemOpen
  \bibfield  {author} {\bibinfo {author} {\bibfnamefont {N.}~\bibnamefont
  {Murdoch}}, \bibinfo {author} {\bibfnamefont {P.}~\bibnamefont {Sanchez}},
  \bibinfo {author} {\bibfnamefont {S.~R.}\ \bibnamefont {Schwartz}}, \ and\
  \bibinfo {author} {\bibfnamefont {H.}~\bibnamefont {Miyamoto}},\ }\enquote
  {\bibinfo {title} {Asteroids iv},}\ \ (\bibinfo  {publisher} {University of
  Arizona Press Space Science Series},\ \bibinfo {year} {2015})\ Chap.\
  \bibinfo {chapter} {Asteroid Surface Geophysics}\BibitemShut {NoStop}%
\bibitem [{\citenamefont {Hofmeister}, \citenamefont {Blum},\ and\
  \citenamefont {Hei{\ss}elmann}(2009)}]{hofmeister2009flow}%
  \BibitemOpen
  \bibfield  {author} {\bibinfo {author} {\bibfnamefont {P.~G.}\ \bibnamefont
  {Hofmeister}}, \bibinfo {author} {\bibfnamefont {J.}~\bibnamefont {Blum}}, \
  and\ \bibinfo {author} {\bibfnamefont {D.}~\bibnamefont {Hei{\ss}elmann}}\
  }(\bibinfo  {publisher} {AIP Conference Proceedings},\ \bibinfo {year}
  {2009})\BibitemShut {NoStop}%
\bibitem [{\citenamefont {Murdoch}\ \emph
  {et~al.}(2013{\natexlab{a}})\citenamefont {Murdoch}, \citenamefont {Rozitis},
  \citenamefont {Green}, \citenamefont {de~Lophem}, \citenamefont {Michel},\
  and\ \citenamefont {Losert}}]{murdoch2013A}%
  \BibitemOpen
  \bibfield  {author} {\bibinfo {author} {\bibfnamefont {N.}~\bibnamefont
  {Murdoch}}, \bibinfo {author} {\bibfnamefont {B.}~\bibnamefont {Rozitis}},
  \bibinfo {author} {\bibfnamefont {S.~F.}\ \bibnamefont {Green}}, \bibinfo
  {author} {\bibfnamefont {T.}~\bibnamefont {de~Lophem}}, \bibinfo {author}
  {\bibfnamefont {P.}~\bibnamefont {Michel}}, \ and\ \bibinfo {author}
  {\bibfnamefont {W.}~\bibnamefont {Losert}},\ }\href@noop {} {\bibfield
  {journal} {\bibinfo  {journal} {Granular Matter}\ }\textbf {\bibinfo {volume}
  {15}},\ \bibinfo {pages} {129} (\bibinfo {year}
  {2013}{\natexlab{a}})}\BibitemShut {NoStop}%
\bibitem [{\citenamefont {Murdoch}\ \emph
  {et~al.}(2013{\natexlab{b}})\citenamefont {Murdoch}, \citenamefont {Rozitis},
  \citenamefont {Nordstrom}, \citenamefont {Green}, \citenamefont {Michel},
  \citenamefont {de~Lophem},\ and\ \citenamefont {Losert}}]{murdoch2013B}%
  \BibitemOpen
  \bibfield  {author} {\bibinfo {author} {\bibfnamefont {N.}~\bibnamefont
  {Murdoch}}, \bibinfo {author} {\bibfnamefont {B.}~\bibnamefont {Rozitis}},
  \bibinfo {author} {\bibfnamefont {K.}~\bibnamefont {Nordstrom}}, \bibinfo
  {author} {\bibfnamefont {S.~F.}\ \bibnamefont {Green}}, \bibinfo {author}
  {\bibfnamefont {P.}~\bibnamefont {Michel}}, \bibinfo {author} {\bibfnamefont
  {T.}~\bibnamefont {de~Lophem}}, \ and\ \bibinfo {author} {\bibfnamefont
  {W.}~\bibnamefont {Losert}},\ }\href@noop {} {\bibfield  {journal} {\bibinfo
  {journal} {Physical Review Letters}\ }\textbf {\bibinfo {volume} {110}},\
  \bibinfo {pages} {018307} (\bibinfo {year} {2013}{\natexlab{b}})}\BibitemShut
  {NoStop}%
\bibitem [{\citenamefont {G{\"u}ttler}\ \emph {et~al.}(2013)\citenamefont
  {G{\"u}ttler}, \citenamefont {von Borstel}, \citenamefont {Schr{\"a}pler},\
  and\ \citenamefont {Blum}}]{guttler2013}%
  \BibitemOpen
  \bibfield  {author} {\bibinfo {author} {\bibfnamefont {C.}~\bibnamefont
  {G{\"u}ttler}}, \bibinfo {author} {\bibfnamefont {I.}~\bibnamefont {von
  Borstel}}, \bibinfo {author} {\bibfnamefont {R.}~\bibnamefont
  {Schr{\"a}pler}}, \ and\ \bibinfo {author} {\bibfnamefont {J.}~\bibnamefont
  {Blum}},\ }\href@noop {} {\bibfield  {journal} {\bibinfo  {journal} {Physical
  Review E}\ }\textbf {\bibinfo {volume} {87}},\ \bibinfo {pages} {044201}
  (\bibinfo {year} {2013})}\BibitemShut {NoStop}%
\bibitem [{\citenamefont {Dove}\ and\ \citenamefont
  {Colwell}(2013)}]{dove2013}%
  \BibitemOpen
  \bibfield  {author} {\bibinfo {author} {\bibfnamefont {A.}~\bibnamefont
  {Dove}}\ and\ \bibinfo {author} {\bibfnamefont {J.~E.}\ \bibnamefont
  {Colwell}},\ }in\ \href@noop {} {\emph {\bibinfo {booktitle} {AGU Fall
  Meeting Abstracts}}},\ Vol.~\bibinfo {volume} {1}\ (\bibinfo {year} {2013})\
  p.~\bibinfo {pages} {01}\BibitemShut {NoStop}%
\bibitem [{\citenamefont {Colwell}\ \emph {et~al.}(2015)\citenamefont
  {Colwell}, \citenamefont {Brisset}, \citenamefont {Dove}, \citenamefont
  {Whizin}, \citenamefont {Nagler},\ and\ \citenamefont {Brown}}]{Colwell2015}%
  \BibitemOpen
  \bibfield  {author} {\bibinfo {author} {\bibfnamefont {J.}~\bibnamefont
  {Colwell}}, \bibinfo {author} {\bibfnamefont {J.}~\bibnamefont {Brisset}},
  \bibinfo {author} {\bibfnamefont {A.}~\bibnamefont {Dove}}, \bibinfo {author}
  {\bibfnamefont {A.}~\bibnamefont {Whizin}}, \bibinfo {author} {\bibfnamefont
  {H.}~\bibnamefont {Nagler}}, \ and\ \bibinfo {author} {\bibfnamefont
  {N.}~\bibnamefont {Brown}}\ }(\bibinfo  {publisher} {European Planetary
  Science Congress},\ \bibinfo {year} {2015})\BibitemShut {NoStop}%
\bibitem [{\citenamefont {Colwell}\ and\ \citenamefont
  {Taylor}(1999)}]{Colwell1999}%
  \BibitemOpen
  \bibfield  {author} {\bibinfo {author} {\bibfnamefont {J.~E.}\ \bibnamefont
  {Colwell}}\ and\ \bibinfo {author} {\bibfnamefont {M.}~\bibnamefont
  {Taylor}},\ }\href@noop {} {\bibfield  {journal} {\bibinfo  {journal}
  {Icarus}\ }\textbf {\bibinfo {volume} {138}},\ \bibinfo {pages} {241}
  (\bibinfo {year} {1999})}\BibitemShut {NoStop}%
\bibitem [{\citenamefont {Colwell}(2003)}]{Colwell2003}%
  \BibitemOpen
  \bibfield  {author} {\bibinfo {author} {\bibfnamefont {J.~E.}\ \bibnamefont
  {Colwell}},\ }\href@noop {} {\bibfield  {journal} {\bibinfo  {journal}
  {Icarus}\ }\textbf {\bibinfo {volume} {164}},\ \bibinfo {pages} {188}
  (\bibinfo {year} {2003})}\BibitemShut {NoStop}%
\bibitem [{\citenamefont {Goldman}\ and\ \citenamefont
  {Umbanhowar}(2008)}]{Goldman}%
  \BibitemOpen
  \bibfield  {author} {\bibinfo {author} {\bibfnamefont {D.~I.}\ \bibnamefont
  {Goldman}}\ and\ \bibinfo {author} {\bibfnamefont {P.}~\bibnamefont
  {Umbanhowar}},\ }\href@noop {} {\bibfield  {journal} {\bibinfo  {journal}
  {Physics Review E}\ }\textbf {\bibinfo {volume} {77}},\ \bibinfo {pages}
  {021308} (\bibinfo {year} {2008})}\BibitemShut {NoStop}%
\bibitem [{\citenamefont {Altshuler}\ \emph {et~al.}(2013)\citenamefont
  {Altshuler}, \citenamefont {Torres}, \citenamefont {Gonz{\'a}lez-Pita},
  \citenamefont {S{\'a}nchez-Colina}, \citenamefont {P{\'e}rez-Penichet},
  \citenamefont {Waitukaitis},\ and\ \citenamefont {Hidalgo}}]{Altshuler}%
  \BibitemOpen
  \bibfield  {author} {\bibinfo {author} {\bibfnamefont {E.}~\bibnamefont
  {Altshuler}}, \bibinfo {author} {\bibfnamefont {H.}~\bibnamefont {Torres}},
  \bibinfo {author} {\bibfnamefont {A.}~\bibnamefont {Gonz{\'a}lez-Pita}},
  \bibinfo {author} {\bibfnamefont {G.}~\bibnamefont {S{\'a}nchez-Colina}},
  \bibinfo {author} {\bibfnamefont {C.}~\bibnamefont {P{\'e}rez-Penichet}},
  \bibinfo {author} {\bibfnamefont {S.}~\bibnamefont {Waitukaitis}}, \ and\
  \bibinfo {author} {\bibfnamefont {R.~C.}\ \bibnamefont {Hidalgo}},\
  }\href@noop {} {\bibfield  {journal} {\bibinfo  {journal} {arXiv:1305.6796}\
  } (\bibinfo {year} {2013})}\BibitemShut {NoStop}%
\bibitem [{\citenamefont {Paton}\ \emph {et~al.}(2015)\citenamefont {Paton},
  \citenamefont {Green}, \citenamefont {Ball}, \citenamefont {Zarnecki},\ and\
  \citenamefont {Harri}}]{Paton2015}%
  \BibitemOpen
  \bibfield  {author} {\bibinfo {author} {\bibfnamefont {M.}~\bibnamefont
  {Paton}}, \bibinfo {author} {\bibfnamefont {S.}~\bibnamefont {Green}},
  \bibinfo {author} {\bibfnamefont {A.}~\bibnamefont {Ball}}, \bibinfo {author}
  {\bibfnamefont {J.}~\bibnamefont {Zarnecki}}, \ and\ \bibinfo {author}
  {\bibfnamefont {A.-M.}\ \bibnamefont {Harri}},\ }\href@noop {} {\bibfield
  {journal} {\bibinfo  {journal} {Advances in Space Research}\ }\textbf
  {\bibinfo {volume} {56}},\ \bibinfo {pages} {1242} (\bibinfo {year}
  {2015})}\BibitemShut {NoStop}%
\bibitem [{\citenamefont {Beitz}\ \emph {et~al.}(2011)\citenamefont {Beitz},
  \citenamefont {G{\"u}ttler}, \citenamefont {Blum}, \citenamefont {Meisner},
  \citenamefont {Teiser},\ and\ \citenamefont {Wurm}}]{Beitz}%
  \BibitemOpen
  \bibfield  {author} {\bibinfo {author} {\bibfnamefont {E.}~\bibnamefont
  {Beitz}}, \bibinfo {author} {\bibfnamefont {C.}~\bibnamefont {G{\"u}ttler}},
  \bibinfo {author} {\bibfnamefont {J.}~\bibnamefont {Blum}}, \bibinfo {author}
  {\bibfnamefont {T.}~\bibnamefont {Meisner}}, \bibinfo {author} {\bibfnamefont
  {J.}~\bibnamefont {Teiser}}, \ and\ \bibinfo {author} {\bibfnamefont
  {G.}~\bibnamefont {Wurm}},\ }\href@noop {} {\bibfield  {journal} {\bibinfo
  {journal} {The Astrophysical Journal}\ }\textbf {\bibinfo {volume} {736}},\
  \bibinfo {pages} {34} (\bibinfo {year} {2011})}\BibitemShut {NoStop}%
\bibitem [{\citenamefont {Schr{\"a}pler}\ \emph {et~al.}(2012)\citenamefont
  {Schr{\"a}pler}, \citenamefont {Blum}, \citenamefont {Seizinger},\ and\
  \citenamefont {Kley}}]{Schrapler}%
  \BibitemOpen
  \bibfield  {author} {\bibinfo {author} {\bibfnamefont {R.}~\bibnamefont
  {Schr{\"a}pler}}, \bibinfo {author} {\bibfnamefont {J.}~\bibnamefont {Blum}},
  \bibinfo {author} {\bibfnamefont {A.}~\bibnamefont {Seizinger}}, \ and\
  \bibinfo {author} {\bibfnamefont {W.}~\bibnamefont {Kley}},\ }\href@noop {}
  {\bibfield  {journal} {\bibinfo  {journal} {The Astrophysical Journal}\
  }\textbf {\bibinfo {volume} {758}},\ \bibinfo {pages} {35} (\bibinfo {year}
  {2012})}\BibitemShut {NoStop}%
\bibitem [{\citenamefont {Hei{\ss}elmann}\ \emph {et~al.}(2010)\citenamefont
  {Hei{\ss}elmann}, \citenamefont {Blum}, \citenamefont {Fraser},\ and\
  \citenamefont {Wolling}}]{Heibelmann}%
  \BibitemOpen
  \bibfield  {author} {\bibinfo {author} {\bibfnamefont {D.}~\bibnamefont
  {Hei{\ss}elmann}}, \bibinfo {author} {\bibfnamefont {J.}~\bibnamefont
  {Blum}}, \bibinfo {author} {\bibfnamefont {H.~J.}\ \bibnamefont {Fraser}}, \
  and\ \bibinfo {author} {\bibfnamefont {K.}~\bibnamefont {Wolling}},\
  }\href@noop {} {\bibfield  {journal} {\bibinfo  {journal} {Icarus}\ }\textbf
  {\bibinfo {volume} {206}},\ \bibinfo {pages} {424} (\bibinfo {year}
  {2010})}\BibitemShut {NoStop}%
\bibitem [{\citenamefont {Scheeres}\ \emph {et~al.}(2010)\citenamefont
  {Scheeres}, \citenamefont {Hartzell}, \citenamefont {S{\'a}nchez},\ and\
  \citenamefont {Swift}}]{Scheeres2010}%
  \BibitemOpen
  \bibfield  {author} {\bibinfo {author} {\bibfnamefont {D.~J.}\ \bibnamefont
  {Scheeres}}, \bibinfo {author} {\bibfnamefont {C.~M.}\ \bibnamefont
  {Hartzell}}, \bibinfo {author} {\bibfnamefont {P.}~\bibnamefont
  {S{\'a}nchez}}, \ and\ \bibinfo {author} {\bibfnamefont {M.}~\bibnamefont
  {Swift}},\ }\href@noop {} {\bibfield  {journal} {\bibinfo  {journal}
  {Icarus}\ }\textbf {\bibinfo {volume} {210}},\ \bibinfo {pages} {968}
  (\bibinfo {year} {2010})}\BibitemShut {NoStop}%
\bibitem [{\citenamefont {Israr}\ \emph {et~al.}(2014)\citenamefont {Israr},
  \citenamefont {Rivallant}, \citenamefont {Bouvet},\ and\ \citenamefont
  {Barrau}}]{israr2014}%
  \BibitemOpen
  \bibfield  {author} {\bibinfo {author} {\bibfnamefont {H.~A.}\ \bibnamefont
  {Israr}}, \bibinfo {author} {\bibfnamefont {S.}~\bibnamefont {Rivallant}},
  \bibinfo {author} {\bibfnamefont {C.}~\bibnamefont {Bouvet}}, \ and\ \bibinfo
  {author} {\bibfnamefont {J.-J.}\ \bibnamefont {Barrau}},\ }\href@noop {}
  {\bibfield  {journal} {\bibinfo  {journal} {Composites Part A: Applied
  Science and Manufacturing}\ }\textbf {\bibinfo {volume} {62}},\ \bibinfo
  {pages} {16} (\bibinfo {year} {2014})}\BibitemShut {NoStop}%
\bibitem [{\citenamefont {Hartmann}(1978)}]{Hartmann1978}%
  \BibitemOpen
  \bibfield  {author} {\bibinfo {author} {\bibfnamefont {W.~K.}\ \bibnamefont
  {Hartmann}},\ }\href@noop {} {\bibfield  {journal} {\bibinfo  {journal}
  {Icarus}\ }\textbf {\bibinfo {volume} {33}},\ \bibinfo {pages} {50} (\bibinfo
  {year} {1978})}\BibitemShut {NoStop}%
\bibitem [{\citenamefont {Murdoch}\ \emph {et~al.}(2016)\citenamefont
  {Murdoch}, \citenamefont {Avila~Martinez}, \citenamefont {Zenou},
  \citenamefont {Sunday}, \citenamefont {Cherrier}, \citenamefont {Cadu},\ and\
  \citenamefont {Gourinat}}]{Murdoch2016}%
  \BibitemOpen
  \bibfield  {author} {\bibinfo {author} {\bibfnamefont {N.}~\bibnamefont
  {Murdoch}}, \bibinfo {author} {\bibfnamefont {I.}~\bibnamefont
  {Avila~Martinez}}, \bibinfo {author} {\bibfnamefont {E.}~\bibnamefont
  {Zenou}}, \bibinfo {author} {\bibfnamefont {C.}~\bibnamefont {Sunday}},
  \bibinfo {author} {\bibfnamefont {O.}~\bibnamefont {Cherrier}}, \bibinfo
  {author} {\bibfnamefont {A.}~\bibnamefont {Cadu}}, \ and\ \bibinfo {author}
  {\bibfnamefont {Y.}~\bibnamefont {Gourinat}},\ }\href@noop {} {\bibfield
  {journal} {\bibinfo  {journal} {In Prep.}\ } (\bibinfo {year}
  {2016})}\BibitemShut {NoStop}%
\bibitem [{\citenamefont {{YEI Technology}}(2014)}]{YEI}%
  \BibitemOpen
  \bibfield  {author} {\bibinfo {author} {\bibnamefont {{YEI Technology}}},\
  }\href@noop {} {\bibfield  {journal} {\bibinfo  {journal} {YEI 3-Space Sensor
  Data-logging Technical Brief}\ } (\bibinfo {year} {2014})}\BibitemShut
  {NoStop}%
\bibitem [{\citenamefont {Nagurka}\ and\ \citenamefont
  {Huang}(2004)}]{Nagurka}%
  \BibitemOpen
  \bibfield  {author} {\bibinfo {author} {\bibfnamefont {M.}~\bibnamefont
  {Nagurka}}\ and\ \bibinfo {author} {\bibfnamefont {S.}~\bibnamefont
  {Huang}},\ }in\ \href@noop {} {\emph {\bibinfo {booktitle} {Proceedings of
  the 2004 American Control Conference}}},\ Vol.~\bibinfo {volume} {1}\
  (\bibinfo  {publisher} {IEEE},\ \bibinfo {year} {2004})\ pp.\ \bibinfo
  {pages} {499--504}\BibitemShut {NoStop}%
\bibitem [{\citenamefont {Krijt}\ \emph {et~al.}(2013)\citenamefont {Krijt},
  \citenamefont {G{\"u}ttler}, \citenamefont {Hei{\ss}elmann}, \citenamefont
  {Dominik},\ and\ \citenamefont {Tielens}}]{Krijt}%
  \BibitemOpen
  \bibfield  {author} {\bibinfo {author} {\bibfnamefont {S.}~\bibnamefont
  {Krijt}}, \bibinfo {author} {\bibfnamefont {C.}~\bibnamefont {G{\"u}ttler}},
  \bibinfo {author} {\bibfnamefont {D.}~\bibnamefont {Hei{\ss}elmann}},
  \bibinfo {author} {\bibfnamefont {C.}~\bibnamefont {Dominik}}, \ and\
  \bibinfo {author} {\bibfnamefont {A.~G. G.~M.}\ \bibnamefont {Tielens}},\
  }\href@noop {} {\bibfield  {journal} {\bibinfo  {journal} {Journal of Physics
  D: Applied Physics}\ }\textbf {\bibinfo {volume} {46}},\ \bibinfo {pages}
  {435303} (\bibinfo {year} {2013})}\BibitemShut {NoStop}%
\bibitem [{\citenamefont {Ambroso}, \citenamefont {Kamien},\ and\ \citenamefont
  {Durian}(2005)}]{Ambroso2005}%
  \BibitemOpen
  \bibfield  {author} {\bibinfo {author} {\bibfnamefont {M.~A.}\ \bibnamefont
  {Ambroso}}, \bibinfo {author} {\bibfnamefont {R.~D.}\ \bibnamefont {Kamien}},
  \ and\ \bibinfo {author} {\bibfnamefont {D.~J.}\ \bibnamefont {Durian}},\
  }\href@noop {} {\bibfield  {journal} {\bibinfo  {journal} {Physical Review
  E}\ }\textbf {\bibinfo {volume} {72}},\ \bibinfo {pages} {041305} (\bibinfo
  {year} {2005})}\BibitemShut {NoStop}%
\bibitem [{\citenamefont {Russell}\ \emph {et~al.}(2012)\citenamefont
  {Russell}, \citenamefont {Raymond}, \citenamefont {Coradini}, \citenamefont
  {McSween}, \citenamefont {Zuber}, \citenamefont {Nathues}, \citenamefont
  {De~Sanctis}, \citenamefont {Jaumann}, \citenamefont {Konopliv},
  \citenamefont {Preusker} \emph {et~al.}}]{Russell2012}%
  \BibitemOpen
  \bibfield  {author} {\bibinfo {author} {\bibfnamefont {C.~T.}\ \bibnamefont
  {Russell}}, \bibinfo {author} {\bibfnamefont {C.~A.}\ \bibnamefont
  {Raymond}}, \bibinfo {author} {\bibfnamefont {A.}~\bibnamefont {Coradini}},
  \bibinfo {author} {\bibfnamefont {H.~Y.}\ \bibnamefont {McSween}}, \bibinfo
  {author} {\bibfnamefont {M.~T.}\ \bibnamefont {Zuber}}, \bibinfo {author}
  {\bibfnamefont {A.}~\bibnamefont {Nathues}}, \bibinfo {author} {\bibfnamefont
  {M.~C.}\ \bibnamefont {De~Sanctis}}, \bibinfo {author} {\bibfnamefont
  {R.}~\bibnamefont {Jaumann}}, \bibinfo {author} {\bibfnamefont {A.~S.}\
  \bibnamefont {Konopliv}}, \bibinfo {author} {\bibfnamefont {F.}~\bibnamefont
  {Preusker}},  \emph {et~al.},\ }\href@noop {} {\bibfield  {journal} {\bibinfo
   {journal} {Science}\ }\textbf {\bibinfo {volume} {336}},\ \bibinfo {pages}
  {684} (\bibinfo {year} {2012})}\BibitemShut {NoStop}%
\bibitem [{\citenamefont {Carry}\ \emph {et~al.}(2008)\citenamefont {Carry},
  \citenamefont {Dumas}, \citenamefont {Fulchignoni}, \citenamefont {Merline},
  \citenamefont {Berthier}, \citenamefont {Hestroffer}, \citenamefont {Fusco},\
  and\ \citenamefont {Tamblyn}}]{Carry2008}%
  \BibitemOpen
  \bibfield  {author} {\bibinfo {author} {\bibfnamefont {B.}~\bibnamefont
  {Carry}}, \bibinfo {author} {\bibfnamefont {C.}~\bibnamefont {Dumas}},
  \bibinfo {author} {\bibfnamefont {M.}~\bibnamefont {Fulchignoni}}, \bibinfo
  {author} {\bibfnamefont {W.~J.}\ \bibnamefont {Merline}}, \bibinfo {author}
  {\bibfnamefont {J.}~\bibnamefont {Berthier}}, \bibinfo {author}
  {\bibfnamefont {D.}~\bibnamefont {Hestroffer}}, \bibinfo {author}
  {\bibfnamefont {T.}~\bibnamefont {Fusco}}, \ and\ \bibinfo {author}
  {\bibfnamefont {P.}~\bibnamefont {Tamblyn}},\ }\href@noop {} {\bibfield
  {journal} {\bibinfo  {journal} {Astronomy \& Astrophysics}\ }\textbf
  {\bibinfo {volume} {478}},\ \bibinfo {pages} {235} (\bibinfo {year}
  {2008})}\BibitemShut {NoStop}%
\bibitem [{\citenamefont {Katsuragi}\ \emph {et~al.}(2016)\citenamefont
  {Katsuragi} \emph {et~al.}}]{Katsuragi2016}%
  \BibitemOpen
  \bibfield  {author} {\bibinfo {author} {\bibfnamefont {H.}~\bibnamefont
  {Katsuragi}} \emph {et~al.},\ }\href@noop {} {\emph {\bibinfo {title}
  {Physics of Soft Impact and Cratering}}}\ (\bibinfo  {publisher} {Springer},\
  \bibinfo {year} {2016})\BibitemShut {NoStop}%
\bibitem [{\citenamefont {Pak}, \citenamefont {Van~Doorn},\ and\ \citenamefont
  {Behringer}(1995)}]{Pak1995}%
  \BibitemOpen
  \bibfield  {author} {\bibinfo {author} {\bibfnamefont {H.~K.}\ \bibnamefont
  {Pak}}, \bibinfo {author} {\bibfnamefont {E.}~\bibnamefont {Van~Doorn}}, \
  and\ \bibinfo {author} {\bibfnamefont {R.~P.}\ \bibnamefont {Behringer}},\
  }\href {\doibase 10.1103/PhysRevLett.74.4643} {\bibfield  {journal} {\bibinfo
   {journal} {Phys. Rev. Lett.}\ }\textbf {\bibinfo {volume} {74}},\ \bibinfo
  {pages} {4643} (\bibinfo {year} {1995})}\BibitemShut {NoStop}%
\bibitem [{\citenamefont {Katsuragi}\ and\ \citenamefont
  {Durian}(2007)}]{katsuragi2007}%
  \BibitemOpen
  \bibfield  {author} {\bibinfo {author} {\bibfnamefont {H.}~\bibnamefont
  {Katsuragi}}\ and\ \bibinfo {author} {\bibfnamefont {D.~J.}\ \bibnamefont
  {Durian}},\ }\href@noop {} {\bibfield  {journal} {\bibinfo  {journal} {Nature
  Physics}\ }\textbf {\bibinfo {volume} {3}},\ \bibinfo {pages} {420} (\bibinfo
  {year} {2007})}\BibitemShut {NoStop}%
\end{thebibliography}%

\end{document}